\documentclass[a4paper,12pt,centertags]{amsart}
\usepackage{mathptmx}
\usepackage[T1]{fontenc}
\usepackage{textcomp}
\usepackage{dsfont}
\usepackage{pgf}
\usepackage[colorlinks]{hyperref}
\usepackage[abbrev,nobysame]{amsrefs}
\DeclareMathAlphabet{\mathpzc}{T1}{pzc}{m}{it}
\newtheorem{theorem}{Theorem}[section]
\newtheorem{lemma}[theorem]{Lemma}
\newtheorem{proposition}[theorem]{Proposition}
\newtheorem{corollary}[theorem]{Corollary}
\newtheorem{assumption}[theorem]{Assumption}
\theoremstyle{definition}
\newtheorem{definition}[theorem]{Definition}
\theoremstyle{remark}
\newtheorem{remark}[theorem]{Remark}
\numberwithin{equation}{section}

\DeclareMathOperator{\Div}{div}
\DeclareMathOperator{\Tr}{Tr}

\DeclareMathOperator{\supp}{Supp}
\newcommand{\e}{\ensuremath{\mathrm{e}}} 
\newcommand{\im}{\mathrm{i}} 
\newcommand{\N}{\mathbf{N}} 
\newcommand{\Z}{\mathbf{Z}} 
\newcommand{\R}{\mathbf{R}} 
\newcommand{\E}{\mathbb{E}} 
\newcommand{\Q}{\mathpzc{Q}} 
\newcommand{\F}{\mathcal{F}} 
\newcommand{\B}{\mathcal{B}} 
\newcommand{\Torus}{{\mathds{T}_\sss{3}}} 
\newcommand{\W}{\mathpzc{W}} 
\newcommand{\V}{\mathpzc{V}} 
\newcommand{\ep}{\varepsilon}
\newcommand{\alo}{{\alpha_0}} 
\newcommand{\epo}{{\varepsilon_0}} 
\newcommand{\Ps}{\mathpzc{P}} 
\newcommand{\er}[1]{\text{\Tiny$(#1)$}} 
\newcommand{\uno}{\mathds{1}} 
\newcommand{\sss}[1]{{\text{\tiny$#1$}}} 
\newcommand{\ssf}[1]{{\text{\footnotesize$#1$}}} 
\newcommand{\mysf}[1]{\textsf{\bfseries\scriptsize[#1]}}
\newcommand{\loc}{\mathrm{loc}} 
\newcommand{\test}{\mathcal{D}^\infty} 
\newcommand{\Enrg}{\mathcal{E}} 
\newcommand{\Gnrg}{\mathcal{G}} 
\newcommand{\Inrt}{\mathcal{J}} 
\newcommand{\IM}{\mathcal{I}}   
\newcommand{\mf}{\mathcal{C}}   
\newcommand{\tu}[1]{\widetilde{u}^\er{#1}}
\newcommand{\stu}{\widetilde{u}}
\newcommand{\stv}{\widetilde{v}}
\newcommand{\tv}[1]{\widetilde{v}^\er{#1}}
\newcommand{\tw}{\widetilde{w}}
\newcommand{\ps}[2]{\langle #1,#2\rangle_H}
\newcommand{\low}{\mathds{P}_K^l}
\newcommand{\high}{\mathds{P}_K^h}
\newcommand{\auno}{\mysf{A1}}
\newcommand{\adue}{\mysf{A2}}
\newcommand{\atre}{\mysf{A3}}
\newcommand{\aqtr}{\mysf{A4}}
\hypersetup{pdftitle={Analysis of equilibrium states of Markov solutions to the 3D Navier-Stokes equations driven by additive noise},
pdfsubject={Analysis of equilibrium states of Markov solutions to the 3D Navier-Stokes equations driven by additive noise},
pdfauthor={M. Romito},
pdfkeywords={stochastic Navier-Stokes equations, martingale problem, Markov selections, invariant measures, ergodicity, energy balance},
pdfcreator={M. Romito}}
\begin{document}
\title[Equilibrium states of Markov solutions to Navier-Stokes]%
      {Analysis of equilibrium states of Markov solutions to the 3D Navier-Stokes equations driven by additive noise}
\author[M. Romito]{Marco Romito}
\address{Dipartimento di Matematica, Universit\`a di Firenze,
         Viale Morgagni 67/a, 50134 Firenze, Italia}
\email{romito@math.unifi.it}
\subjclass[2000]{Primary 76D05; Secondary 60H15, 35Q30, 60H30, 76M35}
\keywords{stochastic Navier-Stokes equations, martingale problem, Markov selections, invariant measures, ergodicity, energy balance}
\date{\today}
\begin{abstract}
We prove that every Markov solution to the three dimensional Navier-Stokes
equation with periodic boundary conditions driven by additive Gaussian
noise is uniquely ergodic. The convergence to the (unique) invariant
measure is exponentially fast.

Moreover, we give a well-posedness criterion for the equations in terms
of invariant measures. We also analyse the energy balance and identify
the term which ensures equality in the balance.
\end{abstract}
\maketitle
\section{Introduction}
The Navier-Stokes equations on the torus with periodic boundary conditions
forced by additive Gaussian noise are a reasonable model for the analysis
of homogeneous isotropic turbulence for an incompressible Newtonian fluid.
\begin{equation}\label{e:realns}
\begin{cases}
\dot u -\nu\Delta u + (u\cdot\nabla)u + \nabla p  = \dot\eta,\\
\Div u=0.
\end{cases}
\end{equation}
The equations share with their deterministic counterpart the well-known problems
of well-posedness. It is reasonable, and possibly useful, to focus on special
classes of solutions, having additional properties.

This paper completes the analysis developed in \cite{FlaRom06},
\cite{FlaRom07a} and \cite{FlaRom07b} (see also \cite{BloFlaRom07}).
In these papers it was proved that it is possible to show the existence
of a Markov process which solves the equations. Moreover, under some
regularity and non-degeneracy assumptions on the covariance of the
driving noise, it has been shown that the associated Markov transition
kernel is continuous in a space $\W$ with a stronger topology (than the
topology of energy, namely $L^2$) for initial conditions in $\W$.

In this paper we show that, under suitable regularity assumptions on the covariance,
\emph{every} Markov solution admits an invariant measure. Moreover, if the noise
is non-degenerate, the invariant measure is unique and the convergence to the
(unique) invariant measure is exponentially fast.

We stress that similar results have been already obtained by Da Prato \& Debussche
\cite{DapDeb03}, Debussche and Odasso \cite{DebOda06} and Odasso \cite{Oda07},
for solutions obtained as limits of spectral Galerkin approximations to \eqref{e:realns},
and constructed via the Kolmogorov equation associated to the diffusion.
The main improvement of our results is that such conclusions are \emph{generically}
valid for \emph{all} Markov solutions and not restricted to solutions limit to the
Galerkin approximations (this would not make any difference whenever the problem
is well-posed, though) and is general enough to be applied to different problems
(see for instance \cite{BloFlaRom07}). Our analysis is essentially based on the
energy balance (see Definition \ref{d:ems} and Remark \ref{r:assol}), and in turn
shows that such balance is the main and crucial ingredient.

It is worth noticing that the uniquely ergodic results hold for any Markov solution,
hence different Markov solutions have their own (unique) invariant measure.
Well-posedness of \eqref{e:realns} would ensure that the invariant measure is unique.
We prove that the latter condition is also sufficient, as if only one invariant
measure is shared among all Markov solutions, then the problem is well-posed.

Finally, we analyse the energy balance for both the process solution to the
equations and the invariant measure. Due to the lack of regularity of trajectories,
the energy balance is indeed an inequality. We identify the missing term and,
under the invariant measure, we relate it to the \emph{energy flux} through
wave-numbers. According to both the physical and mathematical understanding
of the equations, this term \emph{should} be zero.

A non-zero compensating term from one side would invalidate the equations
as a model for phenomenological theories of turbulence, and from the other side
would show that blow-up is typically true. We stress that neither the former
nor the latter statements are proved here.
\subsection{Details on results}
In the rest of the paper we consider the following abstract
version of the stochastic Navier-Stokes equations \eqref{e:realns}
above,
\begin{equation}\label{e:ns}
du + \nu Au + B(u,u) = \Q^\frac12 dW,
\end{equation}
where $A$ is the Laplace operator on the three-dimensional torus $\Torus$
with periodic boundary conditions and $B$ is the projection onto the space
of divergence-free vector fields with finite energy of the Navier-Stokes
non-linearity (see Section \ref{ss:notations} for more details). Moreover,
$W$ is a cylindrical Wiener process on $H$ and $\Q$ is its covariance
operator. We assume that $\Q$ is a symmetric positive operator. We
shall need additional assumptions on the covariance, as the results
contained in the paper holds under slightly different conditions.
Here we gather the different additional assumptions we shall use.
\begin{assumption}\label{a:mainass}
The following assumptions will be used (one at the time) throughout the paper.
\begin{itemize}
\item[{\auno}] $\Q$ has finite trace on $H$.
\item[{\adue}] there is $\alo>0$ such that $A^{\sss{\frac34}+\alo}Q^\frac12$ is a bounded operator on $H$.
\item[{\atre}] there is $\alo>\frac16$ such that $A^{\sss{\frac34}+\alo}Q^\frac12$ is a bounded operator on $H$.
\item[{\aqtr}] there is $\alo>\frac16$ such that $A^{\sss{\frac34}+\alo}Q^\frac12$ is an invertible bounded operator on $H$, with bounded inverse.
\end{itemize}
\end{assumption}
Notice that each of the above conditions implies the following one. We shall
make clear at every stage of the paper which assumption is used.

The first main result of the paper concerns the long time behaviour
of solutions to equations \eqref{e:ns}. We show that \emph{every}
Markov solution is uniquely ergodic and strongly mixing (Theorem
\ref{t:tight} and Corollary \ref{c:ergodic}). Moreover, under
an additional technical condition (see Remark \ref{r:assol})
we prove that the convergence to the (unique) invariant
measure is exponentially fast (Theorem \ref{t:expo}).

We stress that uniqueness of invariant measure is relative to the
Markov solution it arises from. As we do not know if the martingale
problem associated to equations \eqref{e:ns} is well-posed, in principle
there are plenty of Markov solutions, and so plenty of invariant
measures. In Section \ref{s:further} we study a few properties of
the set of invariant measures. In particular, we show the converse
of the above statement, that is if there is only one common invariant
measure for all Markov solutions, then the martingale problem is well
posed (Theorem \ref{t:wellposed}).

We also give some remarks on symmetries for the
invariant measures (such as translations-invariance).
Finally, we analyse the energy inequality (given as
\mysf{M3} and \mysf{M4} in Definition \ref{d:ems},
see also Remark \ref{r:assol}). In particular, we
identify the missing term in the inequality which,
once added, provides the equality. For an invariant
measure $\mu$, we show that
$$
\nu\ep(\mu) + \iota(\mu) = \frac12\sigma^2,
$$
where $\sss{\frac12}\sigma^2$ is the rate of energy
injected by the external force, $\ep(\mu)=\E^\mu[|\nabla x|^2]$
is the \emph{mean rate of energy dissipation} and
$\iota(\mu)$ is the \emph{mean rate of inertial energy
dissipation}. We show also that $\iota(\mu)$ is given
in terms of the \emph{energy flux} through wave-numbers
(see Frisch \cite{Fri95}) as
$$
\iota(\mu)
=\lim_{K\uparrow\infty}\E^\mu\Bigl[\sum_{\substack{l+m=k\\|k|_\infty\leq K,\\ |m|_\infty>K}}(x_m\cdot\overline{x_k})(m\cdot x_l)\Bigr].
$$
\subsection*{Acknowledgements}
The author wishes to thank D.~Bl\"omker and F.~Flandoli
for the several useful conversations and their helpful
comments. The author is also grateful to A.~Debussche
for having pointed out the inequality used in the proof
of Theorem \ref{t:mildRmoments}.
\section{Notations and previous results}
\subsection{Notations}\label{ss:notations}
Let $\Torus=[0,2\pi]^3$ and let $\test$ be the space of infinitely differentiable
vector fields $\varphi:\R^3\to\R^3$ that are divergence-free, periodic and
$$
\int_\Torus\varphi(x)\,dx=0.
$$
We denote by $H$ the closure of $\test$ in the norm of $L^2(\Torus,\R^3)$,
and similarly by $V$ the closure in the norm of $H^1(\Torus,\R^3)$. Let
$D(A)$ be the set of all $u\in H$ such that $\Delta u\in H$ and define
the \emph{Stokes} operator $A:D(A)\to H$ as $Au=-\Delta u$.
By properly identifying dual spaces, we have that
$D(A)\subset V\subset H=H'\subset V'\subset D(A)'$.

The bi-linear operator $B:V\times V\to V'$ is defined as
$$
\langle B(u,v),w\rangle=\sum_{i,j=1}^3\int_\Torus w_i(x)u_j(x)\frac{\partial v_i(x)}{\partial x_j}\,dx
$$
(see Temam \cite{Tem83} for a more detailed account of all
these notations).

Since $A$ is a linear positive and self-adjoint operator with compact inverse,
we can define powers of $A$. We define two hierarchies of spaces related to
the problem, using powers of $A$. The first is given by the following spaces
of \emph{mild} regularity (since they are larger than the space $V$),
\begin{equation}\label{e:mildspace}
\V_\ep=D(A^{\frac14+\ep}),\qquad\ep\in(0,\frac14],
\end{equation}
while the second is given by the following spaces of \emph{strong}
regularity,
\begin{equation}\label{e:strongspace}
\W_\alpha=D(A^{\theta(\alpha)}),\qquad\alpha\in(0,\infty),
\end{equation}
where $\theta$ is defined as
\begin{equation}\label{e:theta}
\theta(\alpha)=\begin{cases}
\frac{\alpha+1}2,&\qquad\alpha\in(0,\frac12),\\
\alpha+\frac14,  &\qquad\alpha\geq\frac12.
\end{cases}
\end{equation}
Notice that for every $\epo$ and $\alo$ as above,
$$
\W_\alo\subset\W_0=V=\V_{\frac14}\subset\V_\epo.
$$
In the proof of most of the results of the paper we shall use repeatedly the following
inequalities.
\begin{lemma}[Temam \cite{Tem83}*{Lemma 2.1, Part I}]\label{l:Btemam}
If $u\in D(A^{\alpha_1})$, $v\in D(A^{\alpha_2})$ and $w\in D(A^{\alpha_3})$, then
there is a constant $c_0=c_0(\alpha_1, \alpha_2,\alpha_3)$ such that
$$
\ps{B(u,v)}{w}\le c_0|A^{\alpha_1}u|\cdot|A^{\alpha_2+\frac12}v|\cdot|A^{\alpha_3}w|,
$$
where $\alpha_i\geq0$ and $\alpha_1+\alpha_2+\alpha_3\geq\frac34$ if $\alpha_i\neq\frac34$
for all $i=1,2,3$, and $\alpha_1+\alpha_2+\alpha_3>\frac34$ otherwise.
\end{lemma}
\begin{lemma}[\cite{FlaRom07a}*{Lemma D.2}]\label{l:Bnostro}
Let $\alpha>0$ and $u$, $v\in D(A^{\theta(\alpha)})$. If $\alpha\neq\ssf{\frac12}$,
there is a constant $C_0=C_0(\alpha)$ such that
$$
|A^{\alpha-\frac14}B(u,v)|_H\leq C_0|A^{\theta(\alpha)}u|_H |A^{\theta(\alpha)}v|_H,
$$
where $\theta$ is the map defined in \eqref{e:theta}.
If $\alpha=\ssf{\frac12}$, then $B$ maps $D(A^{\sss{\frac34}})\times D(A^{\sss{\frac34}})$
continuously to $D(A^{\frac14-\ep})$, for every $\ep>0$.
\end{lemma}
\subsection{Markov solutions to the Navier-Stokes equations}
In this section we recall a few definitions and result from papers
\cite{FlaRom07a} and \cite{FlaRom07b}, with some additional remarks.
\subsubsection{Almost sure super-martingales}
We say that a process $\theta=(\theta_t)_{t\geq0}$ on a probability space
$(\Omega, \mathbb{P}, \F)$, adapted to a filtration $(\F_t)_{t\geq0}$ is
an \emph{a.\ s.\ super-martingale} if it is $\mathbb{P}$-integrable and
there is a set $T\subset(0,\infty)$ of null Lebesgue measure (that we
call the set of \emph{exceptional times} of $\theta$) such that
\begin{equation}\label{e:assupermart}
\E[\theta_t|\F_s]\leq\theta_s,
\end{equation}
for all $s\not\in T$ and all $t>s$.
\begin{lemma}\label{l:dsm}
If $\theta=(\theta_t)_{t\geq0}$ is an a.\ s.\ super-martingale,
then for every $s\geq0$ and every $\varphi\in C^\infty_c(\R)$
with $\varphi\geq0$ and $\supp\varphi\subset[s,\infty)$,
\begin{equation}\label{e:dsm}
\E\Bigl[\int\varphi'(r)\theta_r\,dr\Big|\F_s\Bigr]\geq0.
\end{equation}
\end{lemma}
\begin{proof}
Fix $s\geq0$ and consider a positive smooth map $\varphi$ with
compact support in $[s,\infty)$. By a change of variable,
using the a.\ s.\ super-martingale property,
\begin{align*}
\E\Bigl[\frac1{\ep}\int(\varphi(r)-\varphi(r-\ep))\theta_r\,dr\Big|\F_s\Bigr]
&=\frac1\ep\E\Bigl[\int_s^\infty\!\!\varphi(r)(\theta_r-\theta_{r+\ep})\,dr\Big|\F_s\Bigr]\geq0,
\end{align*}
and in the limit as $\ep\downarrow0$ we get \eqref{e:dsm}.
\end{proof}
It is easy to see that the converse is true (that is, if \eqref{e:dsm}
holds, then the process is an a.\ s.\ super-martingale) under the
assumption that the $\sigma$-fields $\{\F_t:t\geq0\}$ are countably
generated and $\theta$ is lower semi-continuous (see \cite{FlaRom08}).
\subsubsection{Weak martingale solutions}
Let $\Omega=C([0,\infty);D(A)')$, let $\B$ be the Borel $\sigma$-field
on $\Omega$ and let $\xi:\Omega\to D(A)'$ be the canonical process on
$\Omega$ (that is, $\xi_t(\omega)=\omega(t)$). A filtration can be
defined on $\B$ as $\B_t=\sigma(\xi_s:0\leq s\leq t)$.
\begin{definition}\label{d:ems}
Given $\mu_0\in\Pr(H)$, a probability $P$ on $(\Omega,\B)$ is a solution
starting at $\mu_0$ to the martingale problem associated to the Navier-Stokes
equations \eqref{e:ns} if
\begin{itemize}
\item[\mysf{M1}] $P[L_\loc^\infty([0,\infty);H)\cap L^2_\loc([0,\infty);V)]=1$;
\item[\mysf{M2}] for each $\varphi\in\test$ the process $M_t^\varphi$, defined
$P$--a.\ s.\ on $(\Omega,\B)$ as
$$
M_t^\varphi
= \ps{\xi_t-\xi_0}{\varphi}
 +\nu\int_0^t\ps{\xi_s}{A\varphi}\,ds
 -\int_0^t\ps{B(\xi_s,\varphi)}{\xi_s}\,ds
$$
is square integrable and $(M_t^\varphi,\B_t,P)$ is a continuous martingale
with quadratic variation $[M^\varphi]_t=t|\Q^\sss{\frac12}\varphi|^2_H$;
\item[\mysf{M3}] the process $\Enrg^1_t$, defined $P$--a.\ s.\ on $(\Omega,\B)$ as
$$
\Enrg_t^1
= |\xi_t|^2_H
 +2\nu\int_0^t|\xi_s|_V^2\,ds
 -t\Tr[\Q]
$$
is $P$-integrable and $(\Enrg_t^1,\B_t,P)$ is an a.\ s.\ super-martingale;
\item[\mysf{M4}] for each $n\ge2$, the process $\Enrg^n_t$, defined $P$--a.\ s.\ on
$(\Omega,\B)$ as
$$
\Enrg_t^n
= |\xi_t|^{2n}_H
 +2n\nu\int_0^t|\xi_s|^{2n-2}_H|\xi_s|_V^2\,ds
 -n(2n-1)\Tr[\Q]\int_0^t|\xi_s|_H^{2n-2}\,ds
$$
is $P$-integrable and $(\Enrg_t^n,\B_t,P)$ is an a.\ s.\ super-martingale;
\item[\mysf{M5}] $\mu_0$ is the marginal of $P$ at time $t=0$.
\end{itemize}
\end{definition}
\begin{remark}[enhanced martingale solutions]\label{r:assol}
A slightly different approach has been followed in \cite{BloFlaRom07}
to show existence of Markov solution for a different model (an equation
for surface growth driven by space-time white noise), as the energy
balance has been given in terms of an almost sure property. In the
Navier-Stokes setting of this paper the property reads (some equivalent
statements are possible as in \cite{BloFlaRom07})
\begin{itemize}
\item[\mysf{M3-as}] there is a set $T_{P_x}\subset(0,\infty)$ of null
Lebesgue measure such that for all $s\not\in T_{P_x}$ and all $t\geq s$,
\begin{equation*}\label{e:M3as}
P_x[\Gnrg_t(v,z)\leq\Gnrg_s(v,z)] = 1,
\end{equation*}
\end{itemize}
where $\Gnrg$ is defined as
$$
\Gnrg_t(v,z)
=\frac12|v_t|_H^2 + \nu\int_0^t|v_r|_V^2\,dr+\int_0^t\ps{v_r}{B(v_r+z_r,z_r)}\,dr,
$$
$z$ is the solution to the Stokes problem \eqref{e:stokes} and $v=\xi-z$.
It is possible to show that, as in \cite{BloFlaRom07}, there exist Markov
solutions which additionally satisfy \mysf{M3-as}. We shall assume this
statement (see \cite{FlaRom08} for more details).
\end{remark}
\subsection{Previous results}\label{ss:previous}
In the next theorems we summarise some results on existence and
regularity of Markov solutions to the Navier-Stokes equations
\eqref{e:ns}. First we show that there is a Markov solution
to the Navier-Stokes equations \eqref{e:ns}.
\begin{theorem}[\cite{FlaRom07a}*{Theorem $4.1$}]\label{t:markov}
Under condition {\auno} of Assumption \ref{a:mainass}, there exists
a family $(P_x)_{x\in H}$ of weak martingale solutions (as defined
above in Definition \ref{d:ems}), with $P_x$ starting at the measure
concentrated on $x$, for each $x\in H$, and the \emph{almost sure
Markov property} holds. More precisely, for every $x\in H$ there is
a set $T\subset(0,\infty)$ of null Lebesgue measure such that for all
$s\not\in T$, all $t\geq s$ and all bounded measurable $\phi:H\to\R$,
$$
\E^{P_x}[\phi(\xi_t)|\B_s] = \E^{P_{\xi_s}}[\phi(\xi_{t-s})].
$$
\end{theorem}
The map $x\mapsto P_x$ is in principle, from the above result, only
measurable. The regularity of dependence from initial condition
can be significantly improved under stronger assumptions on the noise,
as shown by the theorem below.

If $(P_x)_{x\in H}$ is a Markov solution, the transition semigroup%
\footnote{Notice that, due to the Markov property holding only
almost surely, the family of operators $(\Ps_t)_{t\geq0}$ is
not a semigroup, as the semigroup property holds for almost
every time.}
associated to the solution is defined as
\begin{equation}\label{e:transition}
\Ps_t\varphi(x)=\E^{P_x}[\varphi(\xi_t)],
\qquad x\in H,\ t\geq0,
\end{equation}
for every bounded measurable $\varphi:H\to\R$.
\begin{theorem}[\cite{FlaRom07a}*{Theorem $5.11$}]\label{t:sfeller}
Under condition {\aqtr} of Assumption \ref{a:mainass}, the transition
semigroup $(\Ps_t)_{t\geq0}$ associated to \emph{every} Markov solution
$(P_x)_{x\in H}$ is strong Feller in the topology of $\W_{\alo}$. More
precisely, $\Ps_t\phi\in C_b(\W_{\alo})$ for every $t>0$ and every
bounded measurable $\phi:H\to\R$.
\end{theorem}
The regularity result can be given more explicitly in terms of
quasi-Lipschitz regularity (that is, Lipschitz up to a logarithmic
correction) as in \cite{FlaRom07b}, albeit the estimate given there
holds true only for $\alpha_0=\frac34$ (an extension to all values
of $\alpha_0>\ssf{\frac16}$ can be found in \cite{FlaRom08}).
\section{Existence and uniqueness of the invariant measure}
In this section we prove existence of invariant measures by means
of the classical Krylov-Bogoliubov method. Let $(P_x)_{x\in H}$
be a Markov solution and denote by $(\Ps_t)_{t\geq0}$ its
transition semigroup (see \eqref{e:transition}). Let
$x_0\in H$ and
\begin{equation}\label{e:kribog}
\mu_t=\frac1t\int_0^t\Ps_s^*\delta_{x_0},
\end{equation}
where $\delta_{x_0}$ is the Dirac measure concentrated on $x_0$. It is known
(see for example Da~Prato \& Zabczyk \cite{DapZab92}) that any limit point
of the family of probability measures $(\mu_t)_{t\ge0}$ is an invariant measure
for $(\Ps_t)_{t\ge0}$, provided that the family is tight in the topology
where the transition semigroup is Feller.
\begin{theorem}\label{t:tight}
Assume {\adue} of Assumption \ref{a:mainass}. Let $(P_x)_{x\in H}$ be
any Markov solution to the Navier-Stokes equations (see Theorem \ref{t:markov})
and let $(\Ps_t)_{t\geq0}$ be the associated transition semigroup.
Then the family of probability measures $(\mu_t)_{t\geq1}$ is
tight in $\W_\alo$.
\end{theorem}
The above theorem, together with the strong Feller property ensured by
Theorem \ref{t:sfeller} and Doob's theorem (see Da Prato \& Zabczyk \cite{DapZab96}),
immediately imply the following result.
\begin{corollary}\label{c:ergodic}
Under {\aqtr} of Assumption \ref{a:mainass}, \emph{every} Markov
selection to the Navier-Stokes equations has a unique invariant measure
$\mu_\star$, which is strongly mixing. Moreover, there are $\delta>0$
and $\gamma>0$ (depending only on $\alo$) such that
$$
\E^{\mu_\star}[|A^\delta x|_{\W_\alo}^\gamma]<\infty.
$$
\end{corollary}
The convergence of transition probabilities to the unique invariant
measure can be further improved if, under the same assumptions of
above results, we deal with the \emph{enhanced martingale solutions}
introduced in Remark \ref{r:assol}. This is a technical requirement
that makes the proof of Theorem \ref{t:expo} below simple and,
above all, feasible.
\begin{theorem}\label{t:expo}
Assume {\aqtr} of Assumption \ref{a:mainass} and consider an arbitrary
Markov solution $(P_x)_{x\in H}$ made of enhanced martingale solutions
(see Remark \ref{r:assol}). Let $\mu_\star$ be its unique
invariant measure. Then there are constants $C_{exp}>0$ and $a>0$ 
(independent of the Markov solution and depending only on the data
of problem) such that
$$
\|\Ps_t^*\delta_{x_0}-\mu\|_{\textsf{\Tiny TV}}\leq C_{exp}(1+|x_0|_H^2)\e^{-at},
$$
for all $t>0$ and $x_0\in H$, where $\|\cdot\|_{\textsf{\Tiny TV}}$ is the
total variation distance on measures.
\end{theorem}
\begin{remark}
The proof of Theorem \ref{t:expo} above actually shows a slightly
stronger convergence, namely
$$
\sup_{\|\phi\|_V\leq1}|\Ps_t\phi(x_0)-\int\phi(x)\,\mu(dx)|
\leq C_{exp}(1+|x|_H^2)\e^{-at}
$$
for every $x\in H$ and $t\geq0$, with same constants $C_{exp}$
and $a$, where the norm $\|\cdot\|_V$ is defined on Borel measurable
maps $\phi:H\to\R$ as
$$
\|\phi\|_V=\sup_{x\in H}\frac{|\phi(x)|}{1+|x|_H^2}
$$
(see Goldys \& Maslowski \cite{GolMas05} for details).
\end{remark}
From Theorem $13$ of \cite{FlaRom07b} and again from Theorem $4.2.1$
of Da Prato \& Zabczyk \cite{DapZab96} we also deduce the following result.
\begin{corollary}\label{c:equivalent}
Under the assumptions of previous corollary, let $\mu_1$ and $\mu_2$
be the invariant measures associated to two different Markov selections.
Then the two measures are mutually equivalent.
\end{corollary}
The rest of the section is devoted to the proof of Theorems \ref{t:tight}
and \ref{t:expo}.
\subsection{The proof of Theorem \ref{t:tight}}
We fix a Markov solution $(P_x)_{x\in H}$. Prior to the proof of the
theorem, we show two lemmas on momenta of the solution. The second
lemma is the crucial one.
\begin{lemma}\label{l:hexpo}
Assume {\auno} of Assumption \ref{a:mainass}. Then for every
$x\in H$ and $t\geq0$,
$$
\E^{P_x}[|\xi_t|_H^2]\leq |x|_H^2\e^{-2\nu t}+\frac{\sigma^2}{2\nu}(1-\e^{-2\nu t}).
$$
\end{lemma}
\begin{proof}
The result easily follows from the super-martingale property \mysf{M3},
Poincar\'e inequality and Gronwall's lemma (see for example \cite{Rom01}
for details).
\end{proof}
\begin{lemma}
Assume {\adue} of Assumption \ref{a:mainass}. Then there are $C>0$,
$\delta>0$ and $\gamma>0$ depending only on $\epo$, $\alo$, $\nu$ and
$\sigma^2$ (but not on the Markov solution) such that for $x_0\in H$
and $t\geq1$,
$$
\E^{P_{x_0}}[\frac1t\int_0^t|A^\delta\xi_s|_{\W_\alo}^{2\gamma}\,ds]
\leq C(1+|x_0|_H^2).
$$
\end{lemma}
A slight modification of the argument in the proof below provides
an inequality similar to that of the lemma also for $t<1$.
\begin{proof}
Let $\epo\in(0,\frac14]$ with $\epo<2\alo$. We first prove the statement
of the lemma for $x_0\in\V_\epo$.

Consider values $\delta=\delta(\epo,\alo)$, $\gamma=\gamma(\epo,\alo)$
provided by Theorem \ref{t:mildRmoments}. For every fixed value
$M>0$ we choose $R\geq1+2|x_0|_H^2$, whose value will be given explicitly
later, and we denote by $\ep_R$ the small time where the blow-up
estimate \eqref{e:mildblowup} of Theorem \ref{t:mildweakstrong}
holds true.

Fix $t\geq1$ and $\ep\leq\ep_R$, and let $n_\ep\in\N$ be the largest
integer such that $\ep n_\ep\leq t$. By the Markov property,
\begin{align}\label{e:inpezzi}
\mu_t[|A^\delta x|_{\W_\alo}^2\geq M]
&=   \frac1t\int_0^t P_{x_0}[|A^\delta\xi_s|_{\W_\alo}^2\geq M]\,ds\notag\\
&\leq\frac1t\sum_{k=0}^{n_\ep}\int_{k\ep}^{k\ep+\ep}P_{x_0}[|A^\delta\xi_s|_{\W_\alo}^2\geq M]\,ds\notag\\
&=   \frac1t\sum_{k=0}^{n_\ep}\E^{P_{x_0}}\Bigl[\int_{k\ep}^{k\ep+\ep}P_{\xi_{k\ep}}[|A^\delta\xi'_{s-k\ep}|_{\W_\alo}^2\geq M]\,ds\Bigr]\\
&=   \frac1t\sum_{k=0}^{n_\ep}\E^{P_{x_0}}\Bigl[\int_0^\ep P_{\xi_{k\ep}}[|A^\delta\xi'_s|_{\W_\alo}^2\geq M]\,ds\Bigr].\notag
\end{align}
where $\mu_t$ is the measure defined in \eqref{e:kribog}.
Now, by Theorem \ref{t:mildweakstrong}, for every $x\in\V_\epo$ such that $|x|_{\V_\epo}^2\leq R$,
$$
P_x[|A^\delta\xi_s|_{\W_\alo}^2\geq M]
\leq P_x^\er{\epo,R}[|A^\delta\xi_s|_{\W_\alo}^2\geq M]+P_x[\tau^\er{\epo,R}\leq s]
$$
and so, by using \eqref{e:mildblowup} and Chebychev inequality,
\begin{align*}
\lefteqn{P_{\xi_{k\ep}}[|A^\delta\xi'_s|_{\W_\alo}^2\geq M]\leq}\\
&\quad\leq \bigl( P_{\xi_{k\ep}}^\er{\epo,R}[|A^\delta\xi'_s|_{\W_\alo}^2\geq M]
            +P_{\xi_{k\ep}}[\tau^\er{\epo,R}\leq s]\bigr)\uno_{\{|\xi_{k\ep}|_{\V_\epo}^2\leq R\}}
            +\uno_{\{|\xi_{k\ep}|_{\V_\epo}^2>R\}}\\
&\quad\leq \uno_{\{|\xi_{k\ep}|_{\V_\epo}^2>R\}}
     +\frac1{M^\gamma}\E^{P_{\xi_{k\ep}}^\er{\epo,R}}[|A^\delta\xi'_s|_{\W_\alo}^{2\gamma}]
     +c_1\e^{-c_2\frac{R^2}{\ep_R}}.
\end{align*}
We use the above inequality in \eqref{e:inpezzi} and we apply
Theorem \ref{t:mildRmoments} and the previous lemma,
\begin{align*}
\lefteqn{\mu_t[|A^\delta x|_{\W_\alo}^2\geq M]\leq}\\
&\leq\frac1t\sum_{k=0}^{n_\ep}\E^{P_{x_0}}\Bigl[
        \ep\uno_{\{|\xi_{k\ep}|_{\V_\epo}^2>R\}}
       +\frac1{M^\gamma}\E^{P_{\xi_{k\ep}}^\er{\epo,R}}\bigl[\int_0^\ep|A^\delta\xi'_s|_{\W_\alo}^{2\gamma}\,ds\bigr]
       +c_1\ep\e^{-c_2\frac{R^2}{\ep_R}}\Bigr].\\
&\leq\frac1t\sum_{k=0}^{n_\ep}\Bigl(\ep P_{x_0}[|\xi_{k\ep}|_{\V_\epo}^2>R]
                                     +\frac{C}{M^\gamma}(1+\ep+\E^{P_{x_0}}[|\xi_{k\ep}|_H^2])
                                     +c_1\ep\e^{-c_2\frac{R^2}{\ep_R}}\Bigr).\\
&\leq\frac{\ep}t\sum_{k=0}^{n_\ep} P_{x_0}[|\xi_{k\ep}|_{\V_\epo}^2>R]
    +\frac{C n_\ep}{tM^\gamma}(1+\ep+|x_0|_H^2)
    +c_1\frac{n_\ep\ep}{t}\e^{-c_2\frac{R^2}{\ep_R}}\\
&\leq\frac{\ep}t\sum_{k=0}^{n_\ep} P_{x_0}[|\xi_{k\ep}|_{\V_\epo}^2>R]
    +\frac{C}{\ep M^\gamma}(1+|x_0|_H^2)
    +c_1\e^{-c_2\frac{R^2}{\ep_R}},
\end{align*}
Since all computations above are true for all $\ep\leq\ep_R$, if we integrate
for $\ep\in(\frac12\ep_R,\ep_R)$, we get
\begin{multline*}
\frac{\ep_R}2\mu_t[|A^\delta x|_{\W_\alo}^2\geq M]\leq\\
\leq\frac{\ep_R}t\int_{\frac{\ep_R}2}^{\ep_R}\sum_{k=0}^{n_\ep} P_{x_0}[|\xi_{k\ep}|_{\V_\epo}^2>R]\,d\ep
    +\frac{C\log 2}{M^\gamma}(1+|x_0|_H^2)
    +\frac{c_1\ep_R}2\e^{-c_2\frac{R^2}{\ep_R}}.
\end{multline*}
We use the energy inequality and the previous lemma to estimate the only complicated
term in the inequality above,
\begin{align*}
\frac{\ep_R}t\int_{\frac{\ep_R}2}^{\ep_R}\sum_{k=1}^{n_\ep} P_{x_0}[|\xi_{k\ep}|_{\V_\epo}^2>R]\,d\ep
&\leq\frac{\ep_R}t\sum_{k=1}^{n_R}\int_{\frac{\ep_R}2}^{\ep_R}P_{x_0}[|\xi_{k\ep}|_{\V_\epo}^2>R]\,d\ep\\
&\leq\frac{\ep_R}{tR}\sum_{k=1}^{n_R}\E^{P_{x_0}}\Bigl[\int_{\frac{\ep_R}2}^{\ep_R}|\xi_{k\ep}|_V^2\,d\ep\Bigr]\\
&\leq\frac{\ep_R}{tR}\sum_{k=1}^{n_R}\frac1k\E^{P_{x_0}}\Bigl[\int_{k\frac{\ep_R}2}^{k\ep_R}|\xi_r|_V^2\,dr\Bigr]\\
&\leq\frac{\ep_R}{tR}\sum_{k=1}^{n_R}\frac1k c(1+|x_0|_H^2+k\ep_R)\\
&\leq\frac{c\ep_R}{R}\log\frac{1}{\ep_R}(1+|x_0|_H^2),
\end{align*}
where we have set $n_R=n_{\sss{\frac{\ep_R}2}}$ and $n_\ep\leq n_R$ for all $\ep\in[\frac12\ep_R,\ep_R]$.
Since by \eqref{e:mildblowup} the dependence of $\ep_R$ on $R$ is like $R^{-a}$, for some exponent $a$
depending on $\epo$, we may choose $R$ in such a way that for every $t\geq1$,
$$
\mu_t[|A^\delta x|_{\W_\alo}^2\geq M]
\leq\frac{c}{M^b}\log M.
$$
for a suitable $b>0$. In conclusion, the statement of the lemma is proved
for initial conditions $x_0\in \V_\epo$.

If $x_0\in H$, since for every $s>0$ we know that $\xi_s\in\V_\epo$, $P_{x_0}$-a.\ s.,
then by the Markov property,
\begin{align*}
\E^{P_{x_0}}[\int_s^t|A^\delta\xi_r|_{\W_\alo}^{2\gamma}\,dr]
&=\E^{P_{x_0}}\bigl[\E^{P_{\xi_s}}[\int_0^{t-s}|A^\delta\xi'_r|_{\W_\alo}^{2\gamma}\,dr]\bigr]\\
&\leq C(t-s)\E^{P_{x_0}}[1+|\xi_s|_H^2]\\
&\leq Ct(1+|x_0|_H^2),
\end{align*}
where we have used the previous lemma and this same lemma for initial conditions in $\V_\epo$.
Finally, as $s\downarrow0$, the conclusion follows by the monotone convergence theorem.
\end{proof}
\begin{proof}[Proof of Theorem \ref{t:tight}]
Choose an arbitrary point $x_0\in H$ and consider the sequence
of measures $(\mu_t)_{t\geq1}$ defined by formula \eqref{e:kribog}.
Since
$$
\int|A^\delta x|_{\W_\alo}^{2\gamma}\mu_t(dx)
=\frac1{t}\int_0^t|A^\delta\xi_s|_{\W_\alo}^{2\gamma}\,ds,
$$
where the constants $\delta$ and $\gamma$ are those provided by the
previous lemma, it follows by that same lemma that the sequence of measures
is tight in $\W_\alo$.
\end{proof}
\subsection{The proof of Theorem \ref{t:expo}}
As stated in the statement of the theorem, in this section we work
with the \emph{enhanced martingale solutions} defined in Remark
\ref{r:assol}. It means that the energy balance \mysf{M3-as} is
available for proofs. Prior to the proof, we give a few auxiliary
results, summarised in the following lemmas. In the first one
we show that any solution enters in a ball of small energy with
positive probability.
\begin{lemma}[entrance time in a ball of small energy]\label{l:entrancesmall}
Assume {\atre}. Given $R>0$ and $\delta>0$, there exists
$T_1=T_1(\delta,R)$ such that
$$
\inf_{|x|_H^2\leq R}P(T_1,x,\{y:|y|_H^2\leq\delta\})>0.
$$
\end{lemma}
\begin{proof}
Consider a value $k_1=k_1(\delta)$, to be chosen later, and let
$A=\{\omega:\sup_{t\in[0,T_1]}|z_t|_H^2\leq k_1\}$. We know that
for every $x$, the value $P_x[A]>0$ is constant with respect to
$x$. Since $|\xi_t|_H\leq |z_t|_H+|v_t|_H$, we shall estimate
$v$.

For all $\omega\in A$ such that the inequality in \mysf{M3-as}
(at page \pageref{e:M3as}) holds, we have
\begin{align}\label{e:vestimate}
|v_t|_H^2 - |v_s|_H^2 + 2\nu\int_s^t|v_r|_V^2\,dr
&\leq c\int_s^t|z_r|_V|\xi_r|_V|A^{\ssf{\frac14}}v_r|_H\,dr\notag\\
&\leq c\int_s^t(|z_r|_V|v_r|_H^{\frac12}|v_r|_V^{\frac32} + |z_r|_V^2|v_r|_H^{\frac12}|v_r|_V^{\frac32})\,dr\\
&\leq \nu\int_s^t|v_r|_V^2\,dr + k_2\int_s^t|v_r|_H^2\,dr + k_3(t-s),\notag
\end{align}
where we have set $k_2=c(k_1^4+k_1^{\sss{\frac83}})$ and $k_3=c k_1^{\sss{\frac83}}$.
By the Poincar\'e inequality (the first eigenvalue of the Laplace operator on
the torus $\Torus$ is $1$),
$$
|v_t|_H^2 + (\nu-k_2)\int_s^t|v_r|_H^2\,dr
\leq |v_s|_H^2 + k_3 (t-s),
$$
and Gronwall's lemma ensures that
$$
|v_t|_H^2
\leq |x|_H^2\e^{-(\nu-k_2)t}+\frac{k_3}{\nu-k_2}(1-\e^{-(\nu-k_2)t})
\leq R\e^{-(\nu-k_2)t}+\frac{k_3}{\nu-k_2}.
$$
If we choose $k_1$ and $T_1$ in such a way that
$$
k_1\leq\frac14\delta,
\qquad
k_2<\nu,
\qquad
\frac{k_3}{\nu-k_2}\leq\frac18\delta,
\qquad
R\e^{-(\nu-k_2)T_1}\leq\frac18\delta,
$$
we finally obtain that, if $|x|_H^2\leq R$, then
$P_x[\{|\xi_{T_1}|_H^2\leq\delta\}\cap A]=P_x[A]$.
\end{proof}
The second lemma shows that with positive probability
the dynamics enters into a (sufficiently large) ball
of space $V$.
\begin{lemma}[entrance time in a ball of finite dissipation]\label{l:entrancefinite}
Assume {\atre} from Assumption \ref{a:mainass}. Then there exists $\delta>0$
small enough such that there are $T_2=T_2(\delta)>0$ and $R_2=R_2(\delta)>0$
with
$$
\inf_{|x|_H^2\leq\delta}P(T_2,x,\{y:|y|_V^2\leq R_2\})>0.
$$
\end{lemma}
\begin{proof}
Set $T_2=1$ and let $A=\{\sup_{[0,1]}|A^{\sss{\frac58}}z|_H^2\leq k_1\}$,
with $k_1$ to be chosen later, together with $\delta$.

For all $\omega\in A$ for which the inequality in \mysf{M3-as}
(at page \pageref{e:M3as}) holds, we can proceed as in the
proof of Lemma \ref{l:entrancesmall} to get
$|v_t|_H^2\leq\delta+\frac{k_3}{\nu-k_2}$, with $k_2$
and $k_3$ defined similarly. Using \eqref{e:vestimate},
we get
$$
\int_0^1|v_s|_V^2\,ds
\leq\frac1\nu\Bigl[\delta+k_3+k_2\bigl(\delta+\frac{k_3}{\nu-k_2}\bigr)\Bigr]
:=k_4,
$$
where $k_1$ is small enough so that $k_2<\nu$.

Next, we notice that the set $\{r\in[0,1]:|v_r|_V^2\leq 2k_4\}$
is non-empty (its Lebesgue measure is larger than one half). So
for each $r_0$ in such a set, $|v_{r_0}|_V^2\leq 2k_4$. Since
the energy inequality \mysf{M3-as} holds, for a short time
after $r_0$, $v$ coincides with the unique regular solution.
We shall choose $k_1$ and $\delta$ small enough so that the
short time goes well beyond $1$.

Indeed, using \eqref{l:Btemam} (as in \eqref{e:mildRexist} with
$\epo=\sss{\frac14}$), we get for suitable universal
constants $c_1$ and $c_2$,
$$
\frac{d}{dt}|v|_V^2+2\nu|Av|_H^2
\leq\nu|Av|_H^2 + c_1(|v|_V^6 + |A^\frac58 z|_H^4)
\leq\nu|Av|_H^2 + c_1(|v|_V^6 + c_2k_1^4),
$$
and so, if $\varphi(r)=|v_r|_V^2+k_1^{\sss{\frac43}}$,
we have $\varphi(r_0)\leq 2k_4+k_1^{\sss{\frac43}}$ and
$\dot\varphi\leq c_1\varphi^3$. Now, if we choose
$k_1$ and $\delta$ small enough so that
$$
4c_1(2k_4+k_1^\frac43)^2\leq 1
$$
the solution to the differential inequality of $\varphi$
is finite at least up to time $1+r_0$. In particular,
$\varphi(1)\leq(2c_1)^{-\sss{\frac12}}$ and so by easy
computations,
$$
|\xi_{T_2}|_V^2
= |\xi_1|_V^2
\leq (|v_1|_V+|z_1|_V)^2
\leq 2 k_1+\frac2{\sqrt{2c_1}}.
$$
We choose now the last term on the right-hand side of
the above formula as $R_2$. In conclusion,
$P(T_2,x,\{|y|_V^2\leq R_2)\geq P_x[A]$ and again the
value of $P_x[A]$ is independent of $x$.
\end{proof}
In the last auxiliary lemma we show that the dynamics
enters in a compact subset of $\W_\alo$. This is crucial
since the strong Feller property holds in the topology of
$\W_\alo$ (Theorem \ref{t:sfeller}).
\begin{lemma}[entrance time in a ball of high regularity]\label{l:entrancehigh}
Assume {\atre} from Assumption \ref{a:mainass}. Then there is
$\beta>0$ (depending only on $\alo$) such that for every $R_2>0$
there are a time $T_3=T_3(R_2)>0$ and a constant $C=C(R_2)>0$ and
$$
\inf_{|x|_V^2\leq R_2}P(T_3,x,\{y:|A^\beta y|_{\W_\alo}^2\leq C\})>0.
$$
\end{lemma}
\begin{proof}
Given $R_2>0$, we choose $\beta=\theta''-\theta(\alo)$, $T_3$
and $C$ as given in Lemma \ref{l:return}. Notice that the set
$K=\{y:|A^\beta y|_{\W_\alo}^2\leq C\}$ is a compact
subset of $\W_\alo$.

If $\tau=\tau^\er{\frac14,3 R_2}$ is the time up to which
all solutions starting at $x$ coincide with the unique
solution to problem \eqref{e:mildRns}, then
\begin{align*}
P(T_3,x,K)
&=    P_x[|A^\beta\xi_{T_3}|_{\W_\alo}^2\leq C]\\
&\geq P_x[|A^\beta\xi_{T_3}|_{\W_\alo}^2\leq C,\ \tau>T_3]\\
&\geq P_x^\er{\frac14,3R_2}[|A^\beta\xi_{T_3}|_{\W_\alo}^2\leq C,\ \tau>T_3].
\end{align*}
Now, the conclusion follows from Lemma \ref{l:return}.
\end{proof}
\begin{proof}[Proof of Theorem \ref{t:expo}]
Let $(P_x)_{x\in H}$ be a Markov solution and consider the corresponding
transition kernel $(P(t,x,\cdot))_{t\geq0,x\in H}$. We choose the value
of $\epo$ given in Lemma \ref{l:return} and we consider the value
$\theta''>\theta(\alo)$ provided by the same lemma.

The exponential convergence follows from an abstract result of Goldys \& Maslowski
\cite{GolMas05}*{Theorem 3.1} (which, in turns, is based on results from the book
by Meyn \& Tweedie \cite{MeyTwe93}). More precisely, we need to verify the following
four conditions,
\begin{enumerate}
\item the measures $(P(t,x,\cdot))_{t>0,x\in H}$ are equivalent,
\item $x\to P(t,x,\Gamma)$ is continuous in $\W_\alo$ for all $t>0$ and Borel sets $\Gamma\subset H$,
\item For each $R\geq1$ there are $T_0>0$ and a compact subset $K\subset\W_\alo$ such that
$$
\inf_{|x|_H^2\leq R}P(T_0,x,K)>0,
$$
\item there are $k$, $b$, $c>0$ such that for all $t\geq0$,
$$
\E^{P_x}[|\xi_t|_H^2]\leq k|x|_H^2\e^{-bt}+c.
$$
\end{enumerate}
The first property follows from Theorem $13$ in \cite{FlaRom07b}
(there equivalence is stated only for $x\in\W_\alo$, but it easy to see by the
Markov property that it holds for $x\in H$, as $\W_\alo$ is a set of full measure
for each $P(t,x,\cdot)$). The second property follows from the strong Feller property,
while the fourth property follows from Lemma \ref{l:hexpo}.

We only need to prove the third property. We fix $R\geq1$ and we wish to prove that
there are $T_0=T_0(R)$ and $K=K(R)$ such that
\begin{equation}\label{e:expoclaim}
\inf_{|x|_H^2\leq R}P(T_0,x,\{y:|A^{\theta''} y|_H^2\leq K\})>0.
\end{equation}
We choose the value $\delta$ provided by Lemma \ref{l:entrancefinite}
together with the time $T_2$ and value $R_2$. Corresponding to the
values $R$ and $\delta$, Lemma \ref{l:entrancesmall} gives a time
$T_1$. Moreover, corresponding to $R_2$, Lemma \ref{l:entrancehigh}
provides the time $T_3$ and value $C$.

We set $T_0=T_1+T_2+T_3$, then if $|x|_H^2\leq R$, using three times
the Markov property,
\begin{multline*}
P(T_0,x,K)\geq\\
\geq\inf_{|z|_V^2\leq R_2}P(T_3,z,K)\inf_{|y|_H^2\leq\delta}P(T_2,y,\{|z|_V^2\leq R_2\})\inf_{|x|_H^2\leq R}P(T_1,x,\{|y|_H^2\leq\delta\})
\end{multline*}
and the right-hand side is positive (and bounded from below
independently of $x$) due to Lemma \ref{l:entrancesmall},
\ref{l:entrancefinite} and \ref{l:entrancehigh}.

Finally, the constants $C_{exp}$ and $a$ in the statement of the
theorem are independent of the Markov solution since all computations
either depend on the data (the viscosity $\nu$, the strength of the
noise $\sigma^2$, etc., such as in Lemma \ref{l:hexpo}) or are made
on the regularised problem analysed in the appendix.
\end{proof}
\section{Further analysis of equilibrium states}\label{s:further}
In the previous section we have shown that, under suitable assumptions
on the driving noise, every Markov solution has a unique invariant
measure. As in principle there can be several different Markov
solutions, so are invariant measures.

In the first part of the section we show that well-posedness of the
martingale problem associated to \eqref{e:ns} is equivalent to the
statement that there is only one invariant measure, regardless of
the multiplicity of solutions.

In the second part we give some remarks on symmetries of invariant
measures, while in the third part we analyse the energy balance.
\subsection{A connection between uniqueness of invariant measures and well-posedness of the martingale problem}
\subsubsection{Stationary solutions}
Consider the (unique) invariant measure associated to a Markov
solution $(P_x)_{x\in H}$, as provided by Corollary \ref{c:ergodic},
and define the following probability measure
\begin{equation}\label{e:stat}
P_\star=\int P_x\,\mu_\star(dx)
\end{equation}
\begin{lemma}
The probability measure $P_\star$ defined above is invariant
(in the following, \emph{stationary}) with respect to the time
shifts $\eta_t:\Omega\to\Omega$ defined as
$$
\eta_t(\omega)(s)=\omega(t+s).
$$
\end{lemma}
\begin{proof}
It is sufficient to prove that the finite dimensional marginals of $P_\star$
and $\eta_{s}P_\star$ are the same. The case of one single time is easy,
by invariance of $\mu_\star$.
We consider only the two-dimensional case (one can proceed by induction in
the general case). Consider $t_1<t_2$, then
by the Markov property and invariance of $\mu_\star$,
\begin{align*}
\E^{\eta_{s}P_\star}[f(\xi_{t_1},\xi_{t_2})]
&=\int\E^{P_x}[f(\xi_{s+t_1},\xi_{s+t_2})]\,\mu_\star(dx)\\
&=\int\E^{P_x}\Bigl[\E^{P_{\omega(s+t_1)}}[f(\omega(s+t_1),\xi_{t_2-t_1})]\Bigr]\,\mu_\star(dx)\\
&=\int\E^{P_x}[F(\xi_{s+t_1})]\,\mu_\star(dx)
 =\int\E^{P_x}[F(\xi_{t_1})]\,\mu_\star(dx)\\
&=\int\E^{P_x}[f(\xi_{t_1},\xi_{t_2})]\,\mu_\star(dx)
 =\E^{P_\star}[f(\xi_{t_1},\xi_{t_2})],
\end{align*}
where in the above formula we have set $F(y)=\E^{P_y}[f(y,\xi_{t_2-t_1})]$.
\end{proof}
In turns, the lemma above ensures that $P_\star$ is the unique probability
measure on $\Omega$ such that
\begin{enumerate}
\item $P_\star$ is stationary,
\item $P_\star$ is associated%
\footnote{We say that a probability measure $P$ on $\Omega$ is associated to
a Markov solution $(P_x)_{x\in H}$ if for every $t\geq0$, $P|_{\B_t}^\omega=P_{\omega(t)}$
for $P$-a.\ e.\ $\omega\in\Omega$, where $(P|_{\B_t}^\omega)_{\omega\in\Omega}$ is a
regular conditional probability distribution of $P$ given $\B_t$.}
to the Markov solution $(P_x)_{x\in H}$.
\end{enumerate}
Uniqueness follows easily since $\mu_\star$ is the unique invariant measure
of the Markov solution $(P_x)_{x\in H}$ and since the law of a Markov process
is determined by its one-dimensional (with respect to time) marginal
distributions (as in the proof of Lemma above). We shall see later on that
for a special class of invariant measures this uniqueness statement can
be strengthened (see Proposition~\ref{p:extremaluniq}).

In general one can have several stationary solutions (see for example \cite{Rom01}
for the definition and a different proof of existence) and possibly not all of them
are associated to a Markov solution. Hence we define the two sets,
\begin{gather}
\IM=\{\mu\in\Pr(H):\mu\text{ is the marginal of a stationary solution}\}\label{e:imglobal}\\
\IM_m=\left\{\mu\in\Pr(H):%
  \begin{minipage}{0.6\linewidth}
  $\mu$ is the unique invariant measure associated to a Markov solution
  \end{minipage}%
\right\}\label{e:immarkov}
\end{gather}
and, trivially, $\IM_m\subset\IM$.
\begin{remark}[Topological properties of $\IM$ and $\IM_m$]\label{r:topo}
By the same properties that ensure existence of solutions
(and following similar computations, see for example \cite{FlaRom07a}),
it is easy to see that $\IM$ is
a compact subset of $\Omega$. Moreover, by Corollary
\ref{c:ergodic}, $\IM_m$ and hence $\IM_e$ are relatively
compact in a much stronger topology.
\end{remark}
\subsubsection{A short recap on the selection principle}
It is necessary to give a short account on the procedure which proves
the existence of Markov selection (namely, the proof of Theorem
\ref{t:markov}). We refer to \cite{FlaRom07a}
for all details.

Given $x\in H$, let $\mf(x)\subset\Pr(\Omega)$ be the set of all
weak martingale solutions (according to Definition \ref{d:ems})
to equation \eqref{e:ns}, starting at $x$.

In the proof of Theorem \ref{t:markov} (see \cite{FlaRom07a})
the sets $\mf(x)$ are shrunken to one single element in the
following way. Fix a family $(\lambda_n,f_n)_{n\geq1}$ which
is dense in $[0,\infty)\times C_b(D(A)')$ and consider the
functionals $J_n=J_{\lambda_n,f_n}$, where $J_{\lambda,f}$
is given by
$$
J_{\lambda,f}(P)=\E^P\Bigl[\int_0^\infty\e^{-\lambda t}f(\xi_t)\,dt\Bigr].
$$
for arbitrary $\lambda>0$ and $f:D(A)'\to\R$ upper semi-continuous.
Next, set
$$
\mf_0(x)=\mf(x),
\qquad
\mf_n(x)=\{P\in\mf_{n-1}(x):J_n(P)=\sup_{Q\in\mf_{n-1}(x)}J_n(Q)\}.
$$
All these sets are compact and their intersection is a single element
(the selection associated to this maximised sequence),
$\bigcap_{n\in\N}\mf_n(x)=\{P_x\}$.

Given now a probability measure $\mu$ on $D(A)'$, one can define the set
$\mf(\mu)$ as the set of all probability measures $P$ on $\Omega$ such
that
\begin{enumerate}
\item the marginal at time $0$ of $P$ is $\mu$;
\item there is a a map $x\mapsto Q_x:H\to\Pr(\Omega)$ such that
$P=\int Q_x\,\mu(dx)$ and $Q_x\in\mf(x)$ for all $x$ (in different words,
the conditional distribution of $P$ at time $0$ is made of elements from
sets $(\mf(x))_{x\in H}$).
\end{enumerate}
We can now give the following extension to the selection principle.
\begin{proposition}\label{p:maxim}
Let $(P_x)_{x\in H}$ be the Markov selection associated to the sequence
$(\lambda_n,f_n)_{n\geq1}$. Then the probability $P_\mu=\int_H P_x\,\mu(dx)$
is the unique maximiser associated to the sequence $(\lambda_n,f_n)_{n\geq1}$.
More precisely,
\begin{align*}
J_1(P_\mu)&=\sup_{P\in\mf_0(\mu)}J_1(P),\\
\ldots    &\qquad\ldots,\\
J_n(P_\mu)&=\sup_{P\in\mf_{n-1}(\mu)}J_n(P),\\
\ldots    &\qquad\ldots.
\end{align*}
\end{proposition}
\begin{proof}
Since each $Q\in\mf(\mu)$ is given by $Q=\int Q_x\,\mu(dx)$, for some
$x\mapsto Q_x$, by linearity of the map $J_1$ it easily follows that
$P_\mu\in\mf_1(\mu)$. Moreover, each $Q\in\mf_1(\mu)$ has a similar
structure: $Q=\int Q_x\,\mu(dx)$ and $Q_x\in\mf_1(x)$ for
$\mu$-a.\ e.\ $x\in H$. In fact, $J_1(Q_x)\le J_1(P_x)$, $\mu$-a.\ s., and
$J_1(Q)=J_1(P_\mu)$, and so $J_1(Q_x)=J_1(P_x)$, for $\mu$-a.\ e. $x$. By
induction, $P_\mu\in\mf_n(\mu)$ and for each $Q=\int Q_x\,\mu(dx)\in\mf_n(\mu)$,
$Q_x\in\mf_n(x)$, for $\mu$-a.\ e.\ $x\in H$.

In conclusion, $P_\mu\in\mf_\infty(\mu)=\bigcap\mf_n(\mu)$ and for each
$Q=\int Q_x\,\mu(dx)\in\mf_\infty(\mu)$, $Q_x\in\mf_\infty(x)$, for
$\mu$-a.\ e. $x\in H$. But we know that each $\mf_\infty(x)$
has exactly one element, $P_x$, so that in conclusion the only element
of $\mf_\infty(\mu)$ is $P_\mu$.
\end{proof}
\subsubsection{Connection with well-posedness}
Now, if we are given a sequence $(\lambda_n,f_n)_{n\in\N}$ as above,
the selection principle provides a Markov solution $(P_x)_{x\in H}$.
Corollary \ref{c:ergodic} ensures that this Markov solution has
a unique invariant measure $\mu_\star$. Moreover, from the proposition
above, the stationary solution $P_\star=\int P_x\mu_\star(dx)$ is
the unique sequential maximiser of the sequence $(J_n)_{n\in\N}$
on $\mf(\mu_\star)$. This justifies, in analogy with the definition
of \eqref{e:imglobal} and \eqref{e:immarkov},  the definition of the
following set,
\begin{equation}\label{e:immaxim}
\IM_e=\left\{\mu:%
\begin{minipage}{0.7\linewidth}
$\mu$ is the invariant measure associated to a Markov solution
obtained by the maximisation procedure for some sequence
$(\lambda_n,f_n)_{n\in\N}$
\end{minipage}
\right\},
\end{equation}
and obviously $\IM_e\subset\IM_m\subset\IM$.
\begin{proposition}\label{p:extremaluniq}
If $\mu_\star\in\IM_e$, then the stationary solution $P_\star$
associated to $\mu_\star$ is the unique stationary measure
in $\mf(\mu_\star)$.
\end{proposition}
\begin{proof}
Since $\mu_\star\in\IM_e$, by definition there is a sequence
$(\lambda_n,f_n)_{n\in\N}$ dense in $[0,\infty)\times C_b(D(A)')$
such that $P_\star$ maximises functionals $J_n=J_{\lambda_n,f_n}$
(one after the other, as explained in Proposition \ref{p:maxim}).
Now, if $\widetilde{P}\in\mf(\mu_\star)$ is a stationary solution,
then
\begin{align*}
J_n(\widetilde{P})
&=\E^{\widetilde{P}}\Bigl[\int_0^\infty\e^{-\lambda_n t}f_n(\xi_t)\,dt\Bigr]
=\Bigl(\int f_n(x)\,\mu_\star(dx)\Bigr)\int_0^\infty\e^{-\lambda_n t}\,dt\\
&=\frac{1}{\lambda_n}\int f_n(x)\,\mu_\star(dx),
\end{align*}
and so $J_n(\widetilde{P})=J_n(P_\star)$ for all $n$. By Proposition
\ref{p:maxim}, it follows that $\widetilde{P}=P_\star$.
\end{proof}
If we consider now Markov solutions as those obtained for the Navier-Stokes
equations, namely each of them is strong Feller and irreducible on $\W_\alo$,
the previous result gives immediately a criterion for well-posedness. In few
words, uniqueness of the invariant measures among Markov solutions is
equivalent to well-posedness of the martingale problem.
\begin{corollary}
Assume that every Markov selection is $\W_\alo$--strong Feller
and fully supported on $\W_\alo$. If $(P_x)_{x\in H}$ and $(P_x')_{x\in H}$
are two Markov selections, with $(P_x)_{x\in H}$ coming from a
maximisation procedure, and they have the same invariant
measure, then they coincide on $\W_\alo$.
\end{corollary}
\begin{proof}
Let $P_\star$ and $P_\star'$ be the stationary solutions associated
to the two selections. If the two selections have the same invariant
measure, it follows from the previous theorem that they have the same stationary
solution, that is $P_\star=P_\star'$. It follows from this that
their conditional probability distributions at time $0$ coincide,
$$
P_x=P_x'\qquad\mu_\star-\text{almost surely}.
$$
By $\W_\alo$--strong Feller regularity and irreducibility they coincide
on every $x\in\W_\alo$.
\end{proof}
We summarise the result in the following theorem. It follows easily from
the previous corollary and from the fact that well-posedness of the
martingale problem is equivalent to uniqueness of Markov selections
(see Theorem $12.2.4$ of Stroock \& Varadhan \cite{StrVar79}).
\begin{theorem}\label{t:wellposed}
Under {\aqtr} of Assumption \ref{a:mainass}, assume that the set $\IM_e$
defined in \eqref{e:immaxim} contains only one invariant measure. Then the
martingale problem associated to the Navier-Stokes equations \eqref{e:ns} is
well-posed on $\W_\alo$ (and hence on $\V_\epo$ for all $\epo>0$).
\end{theorem}
\begin{proof}
We only have to prove that, given two Markov solutions $(P_x)_{x\in H}$
and $(P_x')_{x\in H}$, for every $x\in\V_\epo$ we have $P_x=P_x'$.
This statement holds for $x\in\W_\alo$ by the previous corollary.
Fix $\epo>0$ and $x\in\V_\epo$. Choose $R>>|x|_{\V_\epo}^2$,
then, for every bounded continuous $\phi$, by the Markov property,
$$
\Ps_t\phi(x)
= \E^{P_x}[\Ps_{t-\delta}\phi(\xi_{t-\delta})\uno_{\{\tau^\er{\epo,R}>\delta\}}]
 +\E^{P_x}[\Ps_{t-\delta}\phi(\xi_{t-\delta})\uno_{\{\tau^\er{\epo,R}>\delta\}}]
$$
where $\Ps$ is the transition semigroup of $(P_x)_{x\in H}$. The first
term on the right-hand side is independent of the selection, by
the weak-strong uniqueness of Theorem \ref{t:mildweakstrong}, hence
\begin{multline*}
\Ps_t\phi(x)-\Ps_t'\phi(x)=\\
= \E^{P_x}[\Ps_{t-\delta}\phi(\xi_{t-\delta})\uno_{\{\tau^\er{\epo,R}>\delta\}}]
 +\E^{P_x'}[\Ps_{t-\delta}'\phi(\xi_{t-\delta})\uno_{\{\tau^\er{\epo,R}>\delta\}}]
\end{multline*}
and, by the blow-up estimate of Theorem \ref{t:mildweakstrong}, as $\delta\to0$,
we get $\Ps_t\phi(x)=\Ps_t\phi(x)$ for all $\phi$ and all $t$.
\end{proof}
\subsection{Translations-invariance and other symmetries}\label{s:translinv}
In the analysis of homogeneous isotropic turbulence, for which equations
\eqref{e:realns} can be considered a model, it is interesting to consider
equilibrium states invariant with respect to several symmetries (see
for example Frisch \cite{Fri95}).

Here we are interested in solutions which are \emph{translations-invariant}
(in the physical space). For every $a\in\R^3$, define on $\test$ the map
$m_a:\test\to\test$ as
$$
m_a(\varphi)(x)=\varphi(a+x),
\qquad x\in\R^3
$$
for any $\varphi\in\test$. The map obviously extends to $H$ and $D(A^\alpha)$
for each $\alpha$. By composition, it extends to continuous functions on $H$
(or $D(A^\alpha)$ for every $\alpha$) and, by duality, to probability measures
on $H$. It also extends to $\Omega$ as
$$
m_a(\omega)(t)=m_a(\omega(t)),
\qquad t\geq0,\ \omega\in\Omega,
$$
and, by duality, to probability measures on $\Omega$.

A function (or a measure) is \emph{translations-invariant} if it is invariant
under the action of $(m_a)_{a\in\R^3}$. The Navier-Stokes equations are
translations-invariant, so equation \eqref{e:ns} is translations-invariant only
if such is the noise. The driving noise is translations-invariant if and only
if the covariance $\Q$ commutes with all $m_a$. It is easy to verify that
this is equivalent to have homogeneous noise which, in turns, is equivalent
to have that $\Q$ and $A$ commute. So, easy examples of homogeneous noise
compatible with the properties of Assumption \ref{a:mainass} are
$\Q=A^{-\frac32-\alo}$ for any $\alo$ in the correct range.
\begin{proposition}\label{p:translinv}
Assume that $\Q$ and $A$ commute. Then the following properties hold true.
\begin{enumerate}
\item For every $a\in\R^3$, $m_a$ is a one-to-one map on $\IM$, on $\IM_m$ and on $\IM_e$.
\item There is at least one translations-invariant measure in $\IM$.
\end{enumerate}
\end{proposition}
\begin{proof}
We first show that if $P$ is the law of a solution to equations \eqref{e:ns},
then $m_a P$ is also a solution for every $a\in\R^3$. Since for every $a\in\R^3$
the map $m_a$ is an isometry on $H$, the image of a cylindrical Wiener process
on $H$ is again a cylindrical Wiener process. The assumption on $\Q$ ensures
now that the noise term is translations-invariant and so it is easy to check
that all requirements of either Definition \ref{d:ems} or of any definition
of solutions for the stochastic PDE \eqref{e:realns} available in the literature
(see for example Flandoli \& Gatarek \cite{FlaGat95}, we refer also to \cite{Rom01}
as regarding stationary solutions) are verified.

In particular, if $P$ is stationary, then $m_a P$ is again stationary and so
$m_a$ is a one-to-one map on $\IM$. Moreover, since $\IM$ is closed and
convex (see Remark \ref{r:topo}), it follows that there exists
a translations-invariant measure. Indeed, given $\mu\in\IM$, there is
a stationary solution $P_\mu$ whose marginal is $\mu$. Now, the probability
measure
$$
\widetilde{P}
=\frac{1}{|\Torus|}\int_\Torus m_a P_\mu\,da
$$
is again a stationary solution and its marginal is translations-invariant,
as $m_{a+2\pi k}=m_a$ for every $k\in\Z^3$.

We next prove that $m_a$ maps $\IM_m$ one-to-one. Let $\mu_\star\in\IM_m$ and
consider a Markov solution $(P_x)_{x\in H}$ having $\mu_\star$ as one of its
invariant measures. Fix $a\in\R^3$ and set $Q_x=m_a(P_{m_{-a}(x)})$. It is
easy to verify that $(Q_x)_{x\in H}$ is another Markov solution, since
$$
Q_x|^\omega_{\B_t}
=m_a(P_{m_{-a}(x)})|^\omega_{\B_t}
=m_a(P_{m_{-a}(\omega)(t)})
\qquad P_{m_{-a}(x)}-\text{a.\ s.}
$$
Moreover, $m_a(\mu)$ is an invariant measure of $(Q_x)_{x\in H}$.

Finally, in order to show that $m_a$ maps $\IM_e$ one-to-one, we only need
to find a maximising sequence for the solution $(Q_x)_{x\in H}$ defined above.
Let $(\lambda_n,f_n)_{n\in\N}$ be a maximising sequence for $(P_x)_{x\in H}$,
then $(\lambda_n,f_n\circ m_{-a})_{n\in\N}$ is a maximising sequence for
$(Q_x)_{x\in H}$.
\end{proof}
We stress that in the proposition above existence of a translations invariant
equilibrium measure is granted in $\IM$, but we do not know if such a measure
belongs to $\IM_m$.

Notice finally that if problem \eqref{e:ns} is well-posed, it follows easily
that the unique invariant measure must be \emph{translations-invariant}.

Similar conclusions can be found for other symmetries of the torus, such as
isotropy (invariance with respect to rotations, see for example
\cite{FlaGubHaiRom07} where such
symmetries are discussed in view of a connections between homogeneous
turbulence and equations \eqref{e:realns}).
\subsection{The balance of energy}
In the framework of Markov solutions examined in this paper, the balance
of energy corresponds to the a.\ s.\ super-martingale property \mysf{M3}
(and, more generally, of \mysf{M4}) of Definition \ref{d:ems}. As
clarified in \cite{FlaRom07a}, the two facts
\begin{enumerate}
\item the balance holds only for almost every time,
\item the balance is an inequality, rather than an equality,
\end{enumerate}
correspond to a lack of regularity, in time in the first case and in space
in the second, of solutions to the equations \eqref{e:realns}. From the
point of view of the model, such facts translate to a loss of energy
in the balance.

Generally speaking, the problem could be approached by using the
Doob-Meyer decomposition (which may hold even in this case, where
the energy-balance process $\Enrg^1$ is not continuous and the
filtration $(\B_t)_{t\geq0}$ does not satisfy the usual conditions,
see Dellacherie \& Meyer \cite{DelMey82}). We shall follow a
different approach, due to the lack of regularity of trajectories
solutions to the equations. We shall see that the \emph{bounded
variation} term in  the decomposition of $\Enrg^1$ is a distribution
valued process.

Let $a>0$ and define the operator $L_a=\exp(-aA^{\sss{\frac12}})$.
Given a martingale solution $P_x$ starting at some $x\in H$, there is
a Wiener process $(W_t)_{t\geq0}$ such that the canonical process $\xi$
on $\Omega$ solves \eqref{e:ns}. The process $L_a\xi$ under $P_x$ is
regular enough so that we can use the standard stochastic calculus.
Given an arbitrary $\varphi\in C_c^\infty(\R)$, with support in
$[0,\infty)$, It\^o formula gives
\begin{align*}
d[\varphi(t)|L_a\xi_t|_H^2]
&= \varphi'(t)|L_a\xi_t|_H^2
  +2\varphi(t)\ps{L_a\xi_t}{dL_a\xi_t}\\
&=\varphi'(t)|L_a\xi_t|_H^2
  -2\nu\varphi(t)|L_a\xi_t|_V^2
  -2\varphi(t)\ps{L_a\xi_t}{L_a B(\xi_t,\xi_t)}\\
&\quad +2\varphi(t)\ps{L_{2a}\xi_t}{\Q^\frac12\,dW}
       +\varphi(t)\sigma_a^2,
\end{align*}
where $\sigma_a^2=\Tr[\Q L_{2a}]$, and so, by integrating in time, 
\begin{multline*}
2\nu\int\varphi(t)|L_a\xi_t|_V^2\,dt
+2\int\varphi(t)\ps{L_a\xi_t}{L_aB(\xi_t,\xi_t)}\,dt=\\
= \int\varphi'(t)|L_a\xi_t|_H^2\,dt
 +2\int\varphi(t)\ps{L_{2a}\xi_t}{\Q^\frac12\,dW}\,dt
 +\sigma_a^2\int\varphi(t)\,dt,
\end{multline*}
$P_x$-a.\ s. As $a\downarrow0$, the operator $L_a$ approximates
the identity, so that by the regularity of $\xi$ under $P_x$,
\begin{gather*}
2\nu\int\varphi(t)|L_a\xi_t|_V^2\,dt \longrightarrow 2\nu\int\varphi(t)|\xi_t|_V^2\,dt,\\
\int\varphi'(t)|L_a\xi_t|_H^2\,dt \longrightarrow \int\varphi'(t)|\xi_t|_H^2\,dt,\\
2\int\varphi(t)\ps{L_{2a}\xi_t}{\Q^\frac12\,dW}\,dt \longrightarrow 2\int\varphi(t)\ps{\xi_t}{\Q^\frac12\,dW}\,dt,\\
\sigma_a^2\int\varphi(t)\,dt \longrightarrow \sigma^2\int\varphi(t)\,dt,
\end{gather*}
$P_x$-a.\ s.\ and in $L^1(\Omega)$, where $\sigma^2=\Tr[\Q]$.
In conclusion, the limit
\begin{equation}\label{e:inertial}
\int\Inrt_t(\xi)\varphi(t)\,dt:=\lim_{a\downarrow0}\int_0^\infty\varphi(t)\langle L_a\xi_t, L_a B(\xi_t,\xi_t)\rangle\,dt
\end{equation}
exists $P_x$-a.\ s.\ and in $L^1(\Omega)$, and defines a distributions-valued
random variable. Moreover, $\Inrt(\xi)$ depends only on $\xi$ (that is, on $P_x$)
and not on the approximation operators $(L_a)_{a>0}$ used. We finally have
\begin{multline}\label{e:jito}
2\nu\int\varphi(t)|\xi_t|_V^2\,dt
+2\int\varphi(t)\Inrt_t(\xi)\,dt=\\
= \int\varphi'(t)|\xi_t|_H^2\,dt
 +2\int\varphi(t)\ps{\xi_t}{\Q^\frac12\,dW}\,dt
 +\sigma^2\int\varphi(t)\,dt,
\end{multline}
The previous computations and Lemma \ref{l:dsm} provide finally the
following result. In few words, the next theorem states that the
term $\Inrt(\xi)$ plays the role of the increasing process in
the Doob-Meyer decomposition of the a.\ s.\ super-martingale
$(\Enrg^1_t)_{t\geq0}$ (defined by property \mysf{M3} of Definition
\ref{d:ems}).
\begin{theorem}\label{t:doobmeyer}
Given a martingale solution $P_x$, there exists a distribution-valued
random variable $\Inrt(\xi)$, defined by \eqref{e:inertial}, such that
the (distribution-valued) process
$\Enrg^1_t+2\E^{P_x}[\int\Inrt_r(\xi)\,dr|\B_t]$
is a distribution-valued martingale, that is for every $s\geq0$
and every $\varphi\in C_c^\infty(\R)$ with $\supp\varphi\subset[s,\infty)$,
$$
\E^{P_x}\Bigl[\int\varphi'(t)(\Enrg^1_t+2\E^{P_x}\bigl[\int\Inrt_r(\xi)\,dr\big|\B_t\bigr])\,dt\,\Big|\,\B_s\Bigr]
=0.
$$
Moreover, $\Inrt(\xi)$ is a positive distribution, in the sense that
for every $\varphi\in C_c^\infty(\R)$ with $\varphi\geq0$ and
$\supp\varphi\subset[s,\infty)$,
\begin{equation}\label{e:inrtpos}
\E^{P_x}\Bigl[\int\varphi(r)\Inrt_r(\xi)\,dr\Big|\B_s\Bigr]\geq0.
\end{equation}
\end{theorem}
\begin{proof}
The first part of the theorem follows easily, since
\begin{align*}
\int\varphi'(t)\Enrg^1_t\,dt
&=\int\varphi'(t)\Bigl(|\xi_t|_H^2+2\nu\int_0^t|\xi_r|_V^2\,dr-\sigma^2 t\Bigr)\,dt\\
&=\int\varphi'(t)|\xi_t|_H^2\,dt - 2\nu\int\varphi(t)|\xi_t|_V^2\,dt + \sigma^2\int\varphi(t)\,dt,
\end{align*}
and so, using the above computation and formula \eqref{e:jito},
we get the conclusion. The second part is a consequence of
the first part (the martingale property) and the fact that
$(\Enrg_t^1)_{t\geq0}$ is an a.\ s.\ super-martingale.
\end{proof}
\begin{remark}
The It\^o formula applied to $\varphi(t)|L_a\xi|_H^{2n}$
provides an analysis of the a.\ s.\ super-martingale $\Enrg^n$,
defined in property \mysf{M4} of Definition \ref{d:ems}, similar
to that developed above for $\Enrg^1$ and $\Inrt(\xi)$.
\end{remark}
\begin{remark}
Duchon \& Robert \cite{DucRob00} show that the energy equality holds
for suitable weak solutions to the deterministic Navier-Stokes equations
if one takes into account the additional term $\mathcal{D}$, a distribution
in space and time, obtained by means of the limit of space-time
regularisations. Their computations in our setting would lead to
a random distribution $\mathcal{D}(\xi)(t,x)$ in space and time and
$$
\Inrt(\xi)(t) = \langle \mathcal{D}(\xi),\uno_{\Torus}\rangle = \int_{\Torus}\mathcal{D}(\xi)(t,x)\,dx.
$$
This is only formal because in principle our solutions are not
\emph{suitable} (see \cite{Rom01} for existence of suitable
solutions in the stochastic setting).

Moreover, they relate the quantity $\mathcal{D}$ to the four-fifth
law in turbulence theory (see for instance Frisch \cite{Fri95}).
\end{remark}
\subsubsection{The mean rate of inertial energy dissipation}\label{ss:inertial}
Consider a Markov solution $(P_x)_{x\in H}$ and let $\mu_\star$
be its unique invariant measure. Define the \emph{mean rate of
energy dissipation} as
$$
\ep(\mu_\star):=\E^{\mu_\star}[|x|_V^2].
$$
We know that $2\nu\ep(\mu_\star)\leq\sigma^2$. We can as well
consider the expectation with respect to the stationary solution
$P_{\mu_\star}$ of the distribution $\Inrt$ defined in the previous
section. As $\mu_\star$ is an invariant measure, the distribution
$\varphi\longrightarrow \E^{P_{\mu_\star}}[\langle\Inrt(\xi),\varphi\rangle]$
is invariant with respect to time-shifts. Hence there is a constant
$\iota(\mu_\star)$, that we call \emph{mean rate of inertial energy
dissipation}, such that
$$
\E^{P_{\mu_\star}}[\int\Inrt_t(\xi)\varphi(t)\,dt]
=\iota(\mu_\star)\int_0^\infty\varphi(x)\,dx.
$$
We notice that, as a consequence of \eqref{e:inrtpos},
$$
0\leq\iota(\mu_\star).
$$

By taking the expectation in the balance of energy given in \eqref{e:jito},
we finally obtain the following energy equality,
\begin{equation}\label{e:energyequality}
\nu\ep(\mu_\star) + \iota(\mu_\star) = \frac12\sigma^2.
\end{equation}
The quantity $\iota(\mu_\star)$ can be given as the expectation
of \eqref{e:inertial}. Notice that in this case the expectation
in $\mu^\star$ and the limit in \eqref{e:inertial} commute.

We give a different formulation of $\iota(\mu_\star)$ in terms
of Fourier modes. As the definition of $\Inrt$ (and hence of
$\iota$) is independent of the approximation (as long as the
approximating quantities are regular enough, so that all the
computations are correct), we use a ultraviolet cut-off in the
Fourier space. For every threshold $K$, define the projection
$\low$ of $H$ onto low modes as
$$
\low x
=\sum_{|k|_\infty\leq N}x_k\e^{\im k\cdot x},
\qquad\text{for }x=\sum_{k\in\Z^3}x_k\e^{\im k\cdot x},
$$
and the projection onto high modes $\high=I-\low$.
Applying It\^o formula on $\varphi(t)|\low\xi_t|^2$
as in the previous section, taking the expectation with respect
to $P_{\mu_\star}$ and then getting the limit as $K\uparrow\infty$
yields the following representation formula for $\iota(\mu_\star)$,
$$
\iota(\mu_\star)
=\lim_{K\uparrow\infty}\E^{\mu_\star}[\langle\low x, \low B(x,x)\rangle].
$$
Since $x=\low x+\high x$,
\begin{align*}
\langle\low x, \low B(x,x)\rangle
&= \langle\low x, B(x,\low x)\rangle + \langle\low x, B(x,\high x)\rangle\\
&= \langle\low x, B(x,\high x)\rangle
\end{align*}
as $\langle\low x, B(x,\low x)\rangle$ is the sum of a finite number of terms
(so we can use the anti-symmetric property of the non-linear term without
convergence issues). In conclusion,
\begin{align}\label{e:energyflux}
\iota(\mu_\star)
=\lim_{K\uparrow\infty}\E^{\mu_\star}[\langle\low x, B(x,\high x)\rangle]
=\lim_{K\uparrow\infty}\E^{\mu_\star}\Bigl[\sum_{\substack{l+m=k\\|k|_\infty\leq K,\\ |m|_\infty>K}}(x_m\cdot\overline{x_k})(m\cdot x_l)\Bigr]
\end{align}
Following Frisch \cite{Fri95}*{Section 6.2}, the last term we have
obtained in the formula above is the \emph{energy flux} through
wave-number $K$ and represents then energy transferred form the
scales up to $K$ to smaller scales.

From the previous section we know that $\iota(\mu_\star)\geq0$,
this is a consequence of property \mysf{M3} of Definition
\ref{d:ems}. From a mathematical point of view, existence of
invariant measures with $\iota(\mu_\star)>0$ would be an
evidence for loss of regularity and, in turn, for blow-up.
From a physical point of view, the energy flux through
wave-numbers should converge to zero -- hence, again we would
expect $\iota(\mu_\star)=0$ -- as the energy should flow through
modes essentially only in the inertial range (we refer again to
Frisch \cite{Fri95}).
\begin{proposition}
We have
\begin{enumerate}
\item the map $\mu_\star\mapsto\ep(\mu_\star)$ has a smallest element in $\IM$
      (\emph{solution of largest mean inertial dissipation}),
\item if
      $$
      \lim_{R\uparrow\infty}\sup_{\mu\in\IM_m}\E^{\mu}[(|x|_V^2-R)\uno_{\{|x|_V^2>R\}}] = 0,
      $$
      then there is $\mu_\star\in\IM$ such that $\ep(\mu_\star)=\inf_{\mu\in\IM_m}\ep(\mu_\star)$
      (\emph{solution of smallest mean inertial dissipation}).
\end{enumerate}
\end{proposition}
\begin{proof}
The first part follows easily as $\IM$ is compact (see Remark \ref{r:topo})
and $\mu\to\ep(\mu)$ is lower semi-continuous for the topology
with respect to which $\IM$ is compact. As it regards the second part, we know
by Corollary \ref{c:ergodic} that $\IM_m$ is relatively compact on $C_b(V)$.
Hence, if $M$ is the largest value attained by $\ep$ on $\IM_m$ and
$(\mu_n)_{n\in\N}$ is a maximising sequence, say $\ep(\mu_n)\geq M-\frac1n$,
by compactness there is $\mu_\star$ such that, up to a sub-sequence,
$\mu_n\to\mu_\infty$. Now
\begin{align*}
\ep_R(\mu_n)
&=    \E^{\mu_n}[|x|_V^2\uno_{\{|x|_V^2\leq R\}}] + R\mu_n[|x|_V^2>R]\\
&\geq M-\frac1n-\sup_{\mu\in\IM}\E^{\mu}[(|x|_V^2-R)\uno_{\{|x|_V^2> R\}}],
\end{align*}
where $\ep_R(\mu)=\E^{\mu}[|x|_V^2\wedge R]$. As $n\uparrow\infty$, by continuity
$\ep_R(\mu_n)\to\ep_R(\mu_\infty)$, so
$$
\ep(\mu_\infty)
\geq \ep_R(\mu_\infty)
\geq M-\sup_{\mu\in\IM}\int|x|_V^2\uno_{\{|x|_V^2> R\}}\,\mu(dx).
$$
As $R\to\infty$, it follows that $\ep(\mu_\infty)\geq M$, hence $\ep(\mu_\infty)=M$.
\end{proof}
We have not been able yet to prove the condition given in item (\textbf{2})
of previous proposition.

We also remark that such measures of largest and smallest mean inertial dissipation
may not be unique, as both functionals $\ep(\cdot)$ and $\iota(\cdot)$ are
translations-invariant (see Section \ref{s:translinv}).
\appendix
\section{Analysis of the mildly regular approximated problem}\label{s:mildreg}
Let $R\ge 1$ and let $\chi_R:[0,\infty]\to[0,1]$ be a non-increasing
$C^\infty$ function such that $\chi_R\equiv 1$ on $[0,\frac32R]$,
$\chi_R\equiv 0$ on $[2R,\infty)$ and there is $c>0$ such that
$|\chi_R'|\le\frac{c}{R}$ (see the picture).
\begin{figure}[h]
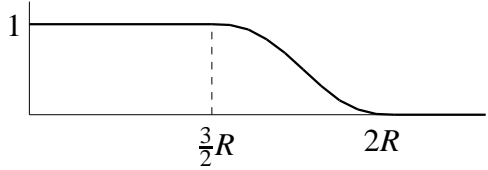

\begin{pgfpicture}{0cm}{-0.2cm}{6cm}{1.5cm}
\pgfsetxvec{\pgfpoint{0.2cm}{0cm}}
\pgfsetyvec{\pgfpoint{0cm}{0.1cm}}
\pgfsetlinewidth{0.01cm}
\pgfline{\pgfxy(0,0)}{\pgfxy(30,0)}
\pgfline{\pgfxy(0,0)}{\pgfxy(0,15)}
\pgfsetlinewidth{0.03cm}
\pgfline{\pgfxy(0,12)}{\pgfxy(12,12)}
\pgfline{\pgfxy(24,0)}{\pgfxy(30,0)}
\pgfline{\pgfxy(12.0, 12.0)}{\pgfxy(13.2, 11.9)}\pgfline{\pgfxy(13.2, 11.9)}{\pgfxy(14.4, 11.3)}
\pgfline{\pgfxy(14.4, 11.3)}{\pgfxy(15.6, 10.0)}\pgfline{\pgfxy(15.6, 10.0)}{\pgfxy(16.8,  8.2)}
\pgfline{\pgfxy(16.8,  8.2)}{\pgfxy(18.0,  6.0)}\pgfline{\pgfxy(18.0,  6.0)}{\pgfxy(19.2,  3.8)}
\pgfline{\pgfxy(19.2,  3.8)}{\pgfxy(20.4,  1.9)}\pgfline{\pgfxy(20.4,  1.9)}{\pgfxy(21.6,  0.7)}
\pgfline{\pgfxy(21.6,  0.7)}{\pgfxy(22.8,  0.1)}\pgfline{\pgfxy(22.8,  0.1)}{\pgfxy(24.0,  0.0)}
\pgfputat{\pgfxy(-1,10.5)}{\pgfbox[center,bottom]{$1$}}
\pgfputat{\pgfxy(11,-4.2)}{\pgfbox[left,center]{$\frac32R$}}
\pgfputat{\pgfxy(22,-3.4)}{\pgfbox[left,center]{$2R$}}
\pgfsetlinewidth{0.01cm}\pgfsetdash{{0.1cm}}{0pt}
\pgfline{\pgfxy(12,0)}{\pgfxy(12,12)}
\end{pgfpicture}
\caption{The cut-off function $\chi_R$}
\end{figure}
\par
Given a value $\epo\in(0,\frac14]$, we consider the following problem in $\V_\epo$,
\begin{equation}\label{e:mildRns}
\begin{cases}
d\tu{\epo,R} + \nu A\tu{\epo,R}+\chi_R(|\tu{\epo,R}|_{\V_\epo}^2)B(\tu{\epo,R},\tu{\epo,R})=\Q^\frac12 dW,\\
\Div\tu{\epo,R}=0.
\end{cases}
\end{equation}
Let $\tau^\er{\epo,R}:\Omega\to[0,\infty)$ be defined as
$$
\tau^\er{\epo,R}(\omega)=\inf\{t\geq0:|\omega(t)|_{\V_{\epo}}^2\ge\frac32R\}
$$
(and $\tau^\er{\epo,R}(\omega)=\infty$ if the above set is empty).
The main aim of this section is to analyse the solutions to the
above problems and their connections to the original Navier-Stokes
equations \eqref{e:ns}.

Before turning to the results on the regularised problem \eqref{e:mildRns},
we remark that in the proof of all results of this section we shall use the
splitting $\tu{\epo,R}=\tv{\epo,R}+z$, where $z$ solves the following linear
Stokes problem
\begin{equation}\label{e:stokes}
\begin{cases}
dz+Az\,dt=\Q^{\frac12}\,dW,\\
z(0)=0,
\end{cases}
\end{equation}
and so $\tv{\epo,R}$ solves the following equation with random coefficients
\begin{equation}\label{e:modns}
\ssf{\frac{d}{dt}}\tv{\epo,R} + \nu A\tv{\epo,R}+\chi_R(|\tu{\epo,R}|_{\V_\epo}^2)B(\tu{\epo,R},\tu{\epo,R})=0.
\end{equation}
\subsection{The weak-strong uniqueness principle}
We first extend the weak-strong uniqueness principle stated
in \cite{FlaRom07a}*{Theorem 5.4} to the above problem
\eqref{e:mildRns}. This is the content of the following
result.
\begin{theorem}\label{t:mildweakstrong}
Assume condition {\adue} of Assumption \ref{a:mainass} and let $\epo\in(0,\frac14]$
with $\epo<2\alo$. Then, for every $x\in\V_\epo$ equation \eqref{e:mildRns} has a
unique martingale solution $P_x^\er{\epo,R}$, with
\begin{equation}\label{e:mildRcont}
P_x^\er{\epo,R}[C([0,\infty);\V_{\epo})]=1.
\end{equation}
Moreover, the following statements hold.
\begin{enumerate}
\item\emph{(weak-strong uniqueness)} On the interval $[0,\tau^\er{\epo,R}]$, the probability
  measure $P_x^\er{\epo,R}$ coincides with any martingale solution $P_x$
  of the original stochastic Navier-Stokes equations \eqref{e:ns}, namely
  $$
  \E^{P_x^\er{\epo,R}}\bigl[\varphi(\xi_t)\uno_{\{\tau^\er{\epo,R}\geq t\}}\bigr]
  =\E^{P_x}\bigl[\varphi(\xi_t)\uno_{\{\tau^\er{\epo,R}\geq t\}}\bigr]
  $$
  for every $t\geq0$ and every bounded measurable $\varphi:H\to\R$.
\item\emph{(blow-up estimate)} There are $c_0>0$, $c_1>0$ and $c_2>0$, depending only
  on $\epo$, such that for every $x\in\V_{\epo}$ with $|x|^2_{\V_\epo}\leq\frac{R}2$,
  \begin{equation}\label{e:mildblowup}
  P_x^\er{\epo,R}[\tau^\er{\epo,R}\leq\delta]
  \leq c_1\e^{-c_2\frac{R^2}{\delta}}
  \end{equation}
  for every $\delta\leq c_0 R^{-\frac1{2\epo}}$.
\end{enumerate}
\end{theorem}
\begin{proof}
The proof is developed in four steps, which are contained in the following
subsections. More precisely, in the first step we prove existence
of solutions for problem \eqref{e:mildRns}, while in the second step
we prove uniqueness. The weak-strong uniqueness principle is then proved
in the third step and the blow-up estimate \eqref{e:mildblowup}
is given in the fourth step.
\medskip\par\noindent\textsl{Step 1: Existence}.
We only show the key estimate for existence. Let $z$ be the solution to the linear
Stokes problem \eqref{e:stokes} and consider $\tv{\epo,R}$ as above.
The usual energy estimate provides (here we use $\stu=\tu{\epo,R}$ and
$\stv=\tv{\epo,R}$ for brevity)
\begin{align}\label{e:mildRexist}
\frac{d}{dt}|\stv|_{\V_\epo}^2+2\nu\|\stv\|_{\V_\epo}^2
&\leq 2\chi_R(|\stu|_{\V_\epo}^2|)\ps{A^{\epo-\frac14}B(\stu, \stu)}{A^{\frac34+\epo}\stv}\notag\\
&\leq c\chi_R(|\stu|_{\V_\epo}^2|)\|\stv\|_{\V_\epo}|A^{\theta(\epo)}\stu|^2\notag\\
&\leq c\chi_R(|\stu|_{\V_\epo}^2|)\|\stv\|_{\V_\epo}(|A^{\theta(\epo)}z|^2+|\stv|_{\V_\epo}^{1+2\epo}\|\stv\|_{\V_\epo}^{1-2\epo})\\
&\leq \nu\|\stv\|_{\V_\epo}^2 + c(|A^{\theta(\epo)}z|^4+|z|_{\V_\epo}^{2+\frac1\epo}+R^{1+\frac1{2\epo}}),\notag
\end{align}
where we have used interpolation inequalities and Lemma \ref{l:Bnostro}, with $\alpha=\epo$.
Notice that, by the choice of $\epo$ with respect to $\alo$, $|A^{\theta(\epo)}z|_H^2$ has
exponential moments.

In order to show \eqref{e:mildRcont}, we show an a-priori estimate for the
derivative in time $\ssf{\frac{d}{dt}}\tv{\epo,R}$ in $L^2(0,T;D(A^{-(\sss{\frac14}-\epo)}))$,
for all $T>0$. The continuity of $\tu{\epo,R}$ then follows from this fact
and continuity of $z$. The a-priori estimate follows by multiplying
the equations by $\ssf{\frac{d}{dt}}A^{2\epo-\sss{\frac12}}\tv{\epo,R}$,
\begin{align*}
|A^{\epo-\frac14}\dot\stv|^2+\frac\nu2\frac{d}{dt}|\stv|_{\V_\epo}^2
&\leq \chi_R(|\stu|_{\V_\epo}^2|)|\ps{A^{\epo-\frac14}B(\stu,\stu)}{A^{\epo-\frac14}\dot\stv}\\
&\leq \frac12|A^{\epo-\frac14}\dot\stv|^2 + c_4\bigl(|A^{\theta(\epo)}z|^4+R^{1+2\epo}\|\stv\|_{\V_\epo}^{2-4\epo}\bigr),
\end{align*}
where we have used the same estimates as in \eqref{e:mildRexist} (and again
$\stu=\tu{\epo,R}$ and $\stv=\tv{\epo,R}$ for brevity).
\medskip\par\noindent\textsl{Step 2: Uniqueness}.
Let $\stu_1$, $\stu_2$ be two solutions of \eqref{e:mildRns} starting at the
same initial condition and set $w=\stu_1-\stu_2$. The new process $w$ solves
the following equation with random coefficients,
\begin{align*}
\dot w + \nu A w
&=\chi_R(|\stu_2|_{\V_\epo}^2)B(\stu_2,\stu_2)-\chi_R(|\stu_1|_{\V_\epo}^2)B(\stu_1,\stu_1)\\
&= \bigl[\chi_R(|\stu_2|_{\V_\epo}^2)-\chi_R(|\stu_1|_{\V_\epo}^2)\bigr]B(\stu_1,\stu_2)
  +\chi_R(|\stu_2|_{\V_\epo}^2)B(w,\stu_2)\\
&\quad  +\chi_R(|\stu_1|_{\V_\epo}^2)B(\stu_1,w)
\end{align*}
and so
\begin{align*}
\frac{d}{dt}|w|_H^2 + 2\nu |w|_V^2
&=2\bigl[\chi_R(|\stu_2|_{\V_\epo}^2)-\chi_R(|\stu_1|_{\V_\epo}^2)\bigr]\ps{w}{B(\stu_1, \stu_2)}\\
&\quad +2\chi_R(|\stu_2|_{\V_\epo}^2)\ps{w}{B(w, \stu_2)}\\
&=\framebox{1}+\framebox{2}.
\end{align*}
Next, we estimate the two terms. In order to estimate the first term,
we first remark that
$$
|\chi_R(|\stu_2|_{\V_\epo}^2)-\chi_R(|\stu_1|_{\V_\epo}^2)|
\leq\frac{c}{\sqrt{R}}|w|_{\V_\epo}[\uno_{[0,2R]}(|\stu_1|^2_{\V_\epo})+\uno_{[0,2R]}(|\stu_2|^2_{\V_\epo})].
$$
By using the above inequality, Lemma \ref{l:Btemam} (with $\alpha_1=\frac12-\epo$,
$\alpha_2=0$ and $\alpha_3=\frac14+\epo$) and interpolation and Young's inequalities,
we get%
\footnote{The inequality has to be slightly modified if $\epo=\frac14$. In such a case
we use Lemma \ref{l:Btemam} with $\alpha_1=\frac12$, $\alpha_2=0$, $\alpha_3=\frac14$,
$$
|\ps{B(\stu_1, \stu_2)}{w}|
\leq c |A^\frac14w|_H |\stu_1|_V |\stu_2|_V.
$$}
\begin{align*}
\framebox{1}
&\leq |\chi_R(|\stu_2|_{\V_\epo}^2)-\chi_R(|\stu_1|_{\V_\epo}^2)\bigr|\,|\ps{B(\stu_1, w)}{\stu_2}|\\
&\leq\frac{c}{\sqrt{R}}\sqrt{2R}|w|_{\V_\epo}|w|_V(|A^{\frac12-\epo}\stu_1|_H+|A^{\frac12-\epo}\stu_2|_H)\\
&\leq c|w|_H^{\frac12-2\epo}|w|_V^{\frac32+2\epo}(|A^{\frac12-\epo}\stu_1|_H+|A^{\frac12-\epo}\stu_2|_H)\\
&\leq\frac\nu2|w|_V^2+c|w|_H^2(|A^{\frac12-\epo}\stu_1|_H+|A^{\frac12-\epo}\stu_2|_H)^{\frac{4}{1-4\epo}}
\end{align*}
For the second term, we use Lemma \ref{l:Btemam} with $\alpha_1=\frac14+\epo$,
$\alpha_2=0$ and $\alpha_3=\frac12-\epo$, and interpolation and Young's inequalities,
\begin{align}\label{e:uniq2term}
\framebox{2}
&=    2\chi_R(|\stu_2|_{\V_\epo}^2)|\ps{\stu_2}{B(w, w)}\notag\\
&\leq 2c\chi_R(|\stu_2|_{\V_\epo}^2)|\stu_2|_{\V_\epo}|w|_V\,|A^{\frac12-\epo}w|_H\notag\\
&\leq 2cR|w|_H^{2\epo}|w|_V^{2(1-\epo)}\\
&\leq \frac12|w|_V^2+cR^{\frac1\epo}|w|_H^2.\notag
\end{align}
Finally, Gronwall's lemma implies that $w\equiv0$, since $w(0)=0$.
\medskip\par\noindent\textsl{Step 3: Weak-strong uniqueness}
The proof works exactly as in \cite{FlaRom07a}*{Theorem 5.12},
we give a short account for the sake of completeness. The proof is
developed in the following steps.
\begin{enumerate}
\item The energy balance of $\tw=\tu{\epo,R}-\xi$, given by
  \begin{align*}
  \widetilde{\Enrg}_t
  &=|\tw(t)|_H^2 + 2\nu\int_0^t|\tw(s)|_V^2\,ds+\\
  &\quad+2\int_0^t\Bigl(\ps{\chi_R(|\tu{\epo,R}|_{\V_\epo}^2)B(\tu{\epo,R},\tu{\epo,R})-B(\xi,\xi)}{\tw}\Bigr)\,ds
  \end{align*}
  is an a.\ s.\ super-martingale under $P_x$.
\item $\tau^\er{\epo,R}$ is a stopping time with respect to the filtration $(\B_t)_{t\geq0}$.
\item The stopped process $(\widetilde{\Enrg}_{t\wedge\tau^\er{\epo,R}})_{t\geq0}$ is again an
      a.\ s.\ super-martingale.
\item The previous step implies the conclusion.
\end{enumerate}
All the above steps can be carried out exactly as in the proof of Theorem $5.12$
of \cite{FlaRom07a} (the key point is that $\tu{\epo,R}$ is continuous in time with
values in $\V_\epo$ with probability one). The only difference is in the last step,
where the estimate of the non-linearity needs to be replaced by the following
estimate,
$$
\ps{\tw}{B(\tu{\epo,R}, \tu{\epo,R})-B(\xi,\xi)}
=-\ps{\tu{\epo,R}}{B(\tw,\tw)}.
$$
Finally, the above estimate can be obtained as in \eqref{e:uniq2term}.
\medskip\par\noindent\textsl{Step 4: The blow-up estimate}
Fix $x\in\V_{\epo}$ with $|x|_{\V_{\epo}}^2\leq\ssf{\frac{R}{2}}$ and $\delta>0$.
Set $\Theta_\delta=\sup_{s\in[0,\delta]}|A^{\theta(\epo)}z|^2$,
then, by slightly modifying inequality \eqref{e:mildRexist}, we get
$$
\frac{d}{dt}|\tv{\epo,R}|_{\V_\epo}^2
\leq c(|A^{\theta(\epo)}z|_H^4+|\tv{\epo,R}|_{\V_\epo}^{2+\frac1\epo})
\leq c(\frac13+\Theta_\delta^2+|\tv{\epo,R}|_{\V_\epo}^2)^{\frac{2\epo+1}{2\epo}},
$$
where $\tv{\epo,R}=\tu{\epo,R}-z$ has been defined in \eqref{e:modns}.
Hence, if we set $\varphi(t)=\frac13+\Theta_\delta^2+|\tv{\epo,R}|_{\V_\epo}^2$,
we end up with a differential inequality that, once solved, gives
$$
\varphi(t)\le\varphi(0)\bigl(1-c\delta\varphi(0)^{\frac1{2\epo}}\bigr)^{-2\epo}
$$
From this, it is easy to show that there is a suitable constant $c_0=c_0(\epo)>0$
such that $|\tu{\epo,R}(s)|_{\V_\epo}^2<\ssf{\frac32}R$ for every $s\leq\delta$, when
$\Theta_\delta\leq\ssf{\frac{R}2}$ and $\delta\leq c_0 R^{-\sss{\frac1{2\epo}}}$.

In conclusion, if $\delta\leq c_0 R^{-\sss{\frac1{2\epo}}}$
and $|x|_{\V_{\epo}}^2\leq\ssf{\frac{R}{2}}$, then
$$
\Theta_\delta\leq\frac{R}2
\quad\Longrightarrow\quad
|\tu{\epo,R}(t)|_{\V_\epo}^2<\frac32 R\quad\text{for }t\leq\delta
\quad\Longrightarrow\quad
\tau^\er{\epo,R}>\delta
$$
and so
$$
P_x^\er{\epo,R}[\tau^\er{\epo,R}\leq\delta]
\leq P_x^\er{\epo,R}[\Theta_\delta>\frac{R}2]
\leq c_1\e^{-c_2\frac{R^2}{\delta}}
$$
with constants $c_1=c_1(\epo)>0$ and $c_2=c_2(\epo)>0$ depending only
on $\epo$ and the last inequality follows easily as in
Proposition $15$ of \cite{FlaRom07b}
(which in turns is a consequence of Proposition $2.16$
of Da Prato \& Zabczyk \cite{DapZab92}).
\end{proof}
\subsection{Moments of norms of stronger regularity}
The proof of the following theorem is based on an inequality taken from
Temam \cite{Tem83}*{Section 4.3, Part I} (see also Odasso \cite{Oda06}). 
\begin{theorem}\label{t:mildRmoments}
Under condition {\adue} of Assumption \ref{a:mainass}, for every
$\epo\in(0,\frac14]$, with $\epo<2\alo$, there are $\delta=\delta(\epo,\alo)>0$
and $\gamma=\gamma(\epo,\alo)>0$ such that
\begin{equation}\label{e:mildRmoment}
\E\Bigl[\int_0^t|A^\delta\tu{\epo,R}|_{\W_\alo}^{2\gamma}\,ds\Bigr]
\leq C[1+t+|x|_H^2],
\end{equation}
where $\tu{\epo,R}$ is the solution to problem \eqref{e:mildRns} starting
at $x\in\V_\epo$ and the value of $C$ is independent of both $x$ and $R$.
\end{theorem}
\begin{proof}
If $\alo\leq\frac14$, we choose $\epo<2\alo$ (such condition is useless for
all other values of $\alo$). The noise is not regular enough to let us work
directly on $\tu{\epo,R}$, so we rely, as in the proof of the previous theorem,
on $\tv{\epo,R}=\tu{\epo,R}-z$. Let $p=\frac12-\epo$ (the value of $p$ could
be slightly improved, but it is beyond our needs) and compute 
$$
\frac{d}{dt}[(1+|\stv|_{\V_{\epo}}^2)^{-p}]
= 2p\frac{\|\stv\|_{\V_{\epo}}^2}{(1+|\stv|_{\V_\epo}^2)^{p+1}}
 -2p\chi_R(|\stu|_{\V_\epo}^2)\frac{\ps{A^{\frac12+2\epo}\stv}{B(\stu,\stu)}}{(1+|\stv|_{\V_\epo}^2)^{p+1}},
$$
where we have set $\stu=\tu{\epo,R}$ and $\stv=\tv{\epo,R}$.
The non-linear term can be estimated as in \eqref{e:mildRexist} to get
$$
\ps{A^{\epo+\frac34}\stv}{A^{\epo-\frac14}B(\stu,\stu)}
\leq \frac12\|\stv\|_{\V_\epo}^2+c\bigl(|A^{\theta(\epo)}z|_H^4+|\stv|_{\V_\epo}^{2+\frac1\epo}\bigr)
$$
and so
$$
\int_0^t\frac{\|\stv\|_{\V_\epo}^2}{(1+|\stv|_{\V_\epo}^2)^{p+1}}\,ds
\leq 1+c\int_0^t|A^{\theta(\epo)}z|^4\,ds+c\int_0^t|\stv|_{\V_\epo}^2\,ds.
$$
The term in $z$ is plain (see for example Da Prato \& Zabzcyk \cite{DapZab92}),
while the term in $|\stv|_{\V_{\epo}}$ can be estimated by means of the energy
inequality \mysf{M3} of Definition \ref{d:ems}. Finally, in order to prove
\eqref{e:mildRmoment}, we use again the energy balance, since by Young's
inequality
$$
\|\stv\|_{\V_\epo}^{2\gamma}
\leq c\Bigl[\frac{\|\stv\|_{\V_\epo}^2}{(1+|\stv|_{\V_\epo}^2)^{p+1}}
    +(1+|\stv|_{\V_\epo}^2)\Bigr]
$$
if $\gamma$ is chosen properly, depending on $p$.

Next, if $\alo>\ssf{\frac14}$, fix $\alpha\in(\ssf{\frac14},\alo)$ and $\epo\in(0,\frac14]$
and choose $p>0$ (whose value will be fixed in dependence of $\alpha$).
We apply It\^o formula to the function $(1+|A^\alpha \tu{\epo,R}(t)|_H^2)^{-p}$,
to get
\begin{align*}
\lefteqn{\frac1{(1+|A^\alpha\stu(t)|_H^2)^p}-\frac1{(1+|A^\alpha x|_H^2)^p}=}\\
&=2p\int_0^t\frac{|A^{\alpha+\frac12}\stu|_H^2}{(1+|A^\alpha\stu|_H^2)^{p+1}}\,ds
+2p\int_0^t\chi_R(|\stu|_{\V_\epo}^2)\frac{\ps{A^{\alpha+\frac12}\stu}{A^{\alpha-\frac12}B(\stu,\stu)}}{(1+|A^\alpha\stu|_H^2)^{p+1}}\,ds\\
&\quad -2p\int_0^t\frac{\ps{A^\alpha\stu}{A^\alpha\Q^\frac12 dW_s}}{(1+|A^\alpha\stu|_H^2)^{p+1}}\,ds
       -p\int_0^t\frac{\sigma_\alpha^2}{(1+|A^\alpha\stu|_H^2)^{p+1}}\,ds\\
&\quad +2p(p+1)\int_0^t\frac{|A^\alpha\Q^\frac12\stu|_H^2}{(1+|A^\alpha\stu|_H^2)^{p+2}}\,ds,
\end{align*}
and we have set again $\stu=\tu{\epo,R}$. The non-linear part is estimated
with Lemma \ref{l:Bnostro}, interpolation and Young's inequalities,
\begin{align*}
\ps{A^{\alpha+\frac12}\stu}{A^{\alpha-\frac12}B(\stu,\stu)}
&\le C|A^{\alpha+\frac12}\stu|_H|A^{\theta(\alpha-\frac14)}\stu|_H^2\\
&\le C|A^{\alpha+\frac12}\stu|_H|A^\alpha\stu|_H^{2\alpha+\frac12}|A^{\alpha+\frac12}\stu|_H^{\frac32-2\alpha}\\
&\le \frac12|A^{\alpha+\frac12}\stu|_H^2+C|A^\alpha\stu|_H^{2\frac{4\alpha+1}{4\alpha-1}}.
\end{align*}
If $\alpha\leq\ssf{\frac14}+\epo$, $\alpha+\ssf{\frac12}>\theta(\alo)$ and
one already knows that some power of $|A^\alpha u|$ has finite moment,
then one can proceed as in the previous case $\alo\leq\ssf{\frac14}$. Otherwise,
as in Temam~\cite{Tem83}, one can iterate the same procedure using
$\alpha-\ssf{\frac12}$ instead of $\alpha$, until the above conditions are
satisfied.
\end{proof}
\subsection{An estimate of the return time to a ball}
The aim of this section is to verify that the probability
of hitting a ball (in a smooth norm) can be uniformly bounded
from below for all initial condition in a given ball.
\begin{lemma}\label{l:return}
Assume condition {\atre} from Assumption \ref{a:mainass}. Then
one can choose $\epo\in(0,\frac14]$ with $\epo<2\alo$ such that
there are $\theta'<\alo+\ssf{\frac12}$, $\theta''>\theta(\alo)$
and a suitable constant $c>0$, and the following statement holds.

For every $R\geq1$ there are values $T_0=T_0(R)$ and $K=K(R)$ such that
for every $x\in\V_\epo$,
\begin{center}
\begin{minipage}{0.43\linewidth}
$$
\mysf{A}\begin{cases}
|x|_{\V_\epo}^2\leq R,\\
\sup_{t\in[0,T_0]}|A^{\theta'}z(t)|_H^2\leq R\\
T_0<c R^{-\frac1{2\epo}},
\end{cases}
$$
\end{minipage}
$\quad\Longrightarrow\quad$
\begin{minipage}{0.4\linewidth}
$$
\mysf{B}\begin{cases}
\tau^\er{\epo,3R}>T_0,\\
|A^{\theta''}\tu{\epo,3R}(T_0)|_H^2\leq K,
\end{cases}
$$
\end{minipage}
\end{center}
where $z$ is the solution to the linear problem \eqref{e:stokes}.
\end{lemma}
\begin{proof}
We choose $\epo=\frac14$ and we set, for brevity, $\stu=\tu{\epo,3R}$ and
$\stv=\stu-z$. The first part of statement \mysf{B} follows as in the proof
of \eqref{e:mildblowup}, if the constant $c$ is chosen accordingly. So, for
every $t\in[0,T_0]$, we know that $|\stu(t)|_V^2\leq 3R$. In particular,
using \eqref{e:mildRexist} and the second statement of \mysf{A}, it follows
that there is a constant $K_0=K_0(R)$ such that
\begin{equation}\label{e:pastbound}
\sup_{[0,T_0]}|\stv(t)|_V^2+\int_0^{T_0}|A\stv(s)|_H^2\,ds\leq K_0(R).
\end{equation}
We next prove the second statement of \mysf{B}.

\textit{Step 1}. We first consider $\alo\in(\ssf{\frac16},\ssf{\frac12}]$ and we choose
$\delta\in(0,\ssf{\frac14})$ so that $\ssf{\frac{\alo}2}<\delta<\delta+\ssf{\frac14}$ (this
condition ensures that $\theta(\alo)<\ssf{\frac12}+\delta$ and
$\theta(\delta+\ssf{\frac14})<\alo+\ssf{\frac12}$). In the case $\alpha=\ssf{\frac12}$
we simply choose a value $\delta\in(\ssf{\frac14},\ssf{\frac12})$.

\textit{Step 2}. \textsl{For all $\omega\in\Omega$ satisfying \mysf{A}, there is
$t_0=t_0(\omega)\in(0,T_0)$ such that $|A\stv(t_0)_H^2\leq 2K_0$}. Indeed, from
\eqref{e:pastbound} it follows that the set $\{t\in(0,T_0):|A\stv(t)_H^2\leq 2K_0\}$
has Lebesgue measure at least $\ssf{\frac{T_0}2}$, and in particular is not empty.

\textit{Step 3}. \textsl{There is $K_1=K_1(R)$ such that for all $\omega\in\Omega$
satisfying \mysf{A}, $|A^{\sss{\frac12}+\delta}\stu(T_0)|_H^2\leq K_1$}. We use
Lemma \ref{l:Bnostro} (with $\alpha=\delta+\ssf{\frac14}$), interpolation of
$D(A^{\theta(\delta+\sss{\frac14})})$ between $V$ and $D(A^{1+\delta})$ and
Young's inequality to obtain the following estimate,
\begin{align*}
\ssf{\frac{d}{dt}}|A^{\frac12+\delta}\stv|_H^2 + 2\nu |A^{1+\delta}\stv|_H^2
&=   2\langle A^{1+2\delta}\stv,B(\stu,\stu)\rangle\\
&\leq c|A^{1+\delta}\stv|_H |A^{\theta(\delta+\sss{\frac14})}\stu|_H^2\\
&\leq \nu |A^{1+\delta}\stv|_H + C\bigl(|A^{\theta(\delta+\sss{\frac14})}z|_H^4 + |\stv|_V^{6+8\delta}\bigr).
\end{align*}
Since $\stv$ is bounded in $V$, the claim easily follows. In the special case
$\alo=\frac12$ one can proceed analogously.

\textit{Step 4}. We choose then $\theta'=\theta(\delta+\ssf{\frac14})$ and
$\theta''=\delta+\ssf{\frac12}$ and the second statement of \mysf{B} follows.

\textit{Step 5}. If $\alo>\frac12$ we iterate the above procedure as in the
proof of Lemma C.1 of \cite{FlaRom07a}, using the two inequalities
\begin{enumerate}
\item[\mysf{i1}] if $m\in\N$, $m\geq1$, there are an integer $p_m$ and $C_m>0$ such that
$$
\ssf{\frac{d}{dt}}|A^{\frac{m}2}\stv|_H^2 + \nu |A^{\frac{m+1}2}\stv|_H^2
\leq C_m(1+|\stv|_V+|A^{\frac{m}2}z|_H)^{p_m},
$$
\item[\mysf{i2}] if $\kappa\geq\frac12$ and $\beta\in[0,\frac12)$, there are $C_{\kappa,\beta}>0$
and $a_\beta>0$ such that
\begin{align*}
\ssf{\frac{d}{dt}}|A^{\kappa+\beta}\stv|_H^2 + \nu |A^{\kappa+\beta+\frac12}\stv|_H^2
&\leq    C_{\kappa,\beta}\bigl[|A^{\kappa+\beta}z|_H^4
       + (|A^{\kappa+\frac12}\stv|_H^2|A^\kappa\stv|^{a_\beta}\\
&\quad + |A^{\kappa+\beta}z|_H^2)|A^{\kappa+\beta}\stv|_H^2\bigr],
\end{align*}
\end{enumerate}
whose proof can be found in the proof of that same lemma.
\end{proof}


\end{document}